\documentclass[aip,jcp,reprint]{revtex4-1}
\usepackage[applemac]{inputenc}
\usepackage{amssymb}
\usepackage{times}
\usepackage{graphicx}
\usepackage{MnSymbol}
\usepackage{subfigure}
\usepackage{natbib}
\usepackage{mciteplus}
\usepackage{dcolumn}% Align table columns on decimal point
\usepackage{bm}% bold math
\usepackage[export]{adjustbox}
\usepackage{mathdots}
\usepackage{multirow}
\usepackage{epstopdf}

\begin{document}

\title{The Uhlenbeck-Ford model: Exact virial coefficients and application as a reference system in fluid-phase free-energy calculations}

\author{Rodolfo Paula Leite}
\affiliation{Instituto de F{\'i}sica Gleb Wataghin, Universidade Estadual de Campinas, UNICAMP, 13083-859 Campinas, S{\~a}o Paulo, Brazil}
\author{Rodrigo Freitas}
\affiliation{Department of Materials Science and Engineering, University of California, Berkeley, CA 94720, U.S.A.}
\author{Rodolfo Azevedo}
\affiliation{Instituto de Computa{\c c}{\~a}o, Universidade Estadual de Campinas, UNICAMP, 13083-852 Campinas, S{\~a}o Paulo Brazil}
\author{Maurice de Koning}
\email{dekoning@ifi.unicamp.br}
\affiliation{Instituto de F{\'i}sica Gleb Wataghin, Universidade Estadual de Campinas, UNICAMP, 13083-859 Campinas, S{\~a}o Paulo, Brazil}

\begin{abstract}
The Uhlenbeck-Ford (UF) model was originally proposed for the theoretical study of imperfect gases, given that all its virial coefficients can be evaluated exactly, in principle. Here, in addition to computing the previously unknown coefficients $B_{11}$ through $B_{13}$, we assess its applicability as a reference system in fluid-phase free-energy calculations using molecular simulation techniques. Our results demonstrate that, although the UF model itself is too soft, appropriately scaled Uhlenbeck-Ford (sUF) models provide robust reference systems that allow accurate fluid-phase free-energy calculations without the need for an intermediate reference model. Indeed, in addition to the accuracy with which their free energies are known and their convenient scaling properties, the fluid is the only thermodynamically stable phase for a wide range of sUF models. This set of favorable properties may potentially put the sUF fluid-phase reference systems on par with the standard role that harmonic and Einstein solids play as reference systems for solid-phase free-energy calculations.
\end{abstract}

\date{\today}

\maketitle

\section{Introduction}
\label{Sec1}

The virial equation of state represents a systematic path to link macroscopic thermodynamic properties of fluid-phase systems to the details of interparticle interactions at the atomic scale.~\cite{Mayer1940,Mason1969,McQuarrie2000,Hansen2006}
It is often expressed as an infinite series for the pressure $P$ in terms of powers of the number density $\rho$,
\begin{equation}
\beta P=\rho + \sum_{n=2}^{\infty} B_n(T) \rho^n,
\label{virial}
\end{equation}
where $\beta \equiv (k_B T)^{-1}$, with $k_B$ Boltzmann's constant and $T$ the absolute temperature, and $B_n(T)$ are the virial coefficients. In general, these coefficients are functions of $T$ and depend on the details of the interparticle interactions of the substance under consideration. For low densities the virial expansion provides a convenient way to quantify deviations with respect to the ideal gas limit of an imperfect fluid. On the other hand, its behavior for higher densities has attracted even more attention in view of the potential link between series convergence properties and fluid-phase thermodynamic stability.~\cite{Groeneveld1962,Baram1990,Baram1991,Clisby2006,Eisenberg2007,Ushcats2012,Bannur2015} Indeed, the possible existence of such a connection has driven the quest to compute higher-order virial coefficients for a number of model systems including the hard-sphere systems,~\cite{Labik2005,Clisby2006,Masters2008,Wheatley2013,Schultz2014} inverse-power-law (IPL) soft-sphere models,~\cite{Wheatley2013,Wheatley2005,Tan2010} and the Lennard-Jones fluid.~\cite{Singh2004,Schultz2009,Schultz2012,Ushcats2014,Feng2015}  

In addition to its relevance in fundamental questions such as those referred to above, the availability of accurate virial equations of states for model systems is also useful in the context of fluid-phase free-energy calculations. Given that free energies are thermal quantities that cannot be expressed in terms of ensemble averages, they are often computed by evaluating the work done along some reversible path between the system of interest and a reference system for which the free energy is known.~\cite{AllenTildesley,Frenkel2002,Chipot2007,Lelievre2010} The choice of a suitable reference depends on the nature of the system of interest. For example, for solid phases the Einstein solid -- a collection of independent harmonic oscillators centered at predefined positions -- provides a natural reference system: its free energy is known analytically, and it is always possible to construct a reversible path to it by choosing the equilibrium positions of the oscillators to coincide with those of the solid of interest.~\cite{Frenkel2002} 

For fluid-phase systems, on the other hand, the most natural reference system is the ideal gas. However, a concern in defining a thermodynamic path to the ideal gas limit is the possible occurrence of a liquid-vapor phase transition along the way, which is usually accompanied by appreciable hysteresis. Although it has been shown that it is in fact possible to obtain accurate liquid-phase free-energy values while traversing an isotherm passing through the two-phase liquid-vapor region,~\cite{Abramo2015} avoiding a coexistence-line crossing is convenient because of the smooth variation of thermodynamic quantities along the way. An often-used approach to achieve this is to carry out the process in two stages.~\cite{Broughton1983I,Broughton1987} First, the interatomic potential of the system of interest is transformed into that of a purely repulsive intermediate reference. Subsequently, given that such systems do not undergo a liquid-vapor transition,~\cite{vanderWaals1873} the excess free-energy with respect to the ideal gas limit can be obtained by means of a reversible second process. However, if an accurate equation of state is available for the intermediate reference system this second step becomes unnecessary, and the absolute free energy of the fluid phase can be determined directly from the work done along the first stage.

In practice, two often-used intermediate repulsive references in the calculation of fluid-phase free energies are the Gaussian-core~\cite{Stillinger1976,Prestipino2005,Prestipino2005a,Ryu2008a} and the IPL~\cite{Hoover1970,Young1984,Prestipino2005,deKoning2001,Michelon2010,Greeff2008,Moriarty2014} soft-sphere models. For the former, no adimensional form for the virial expansion exists, meaning that for each value of $T$ and choice of potential parameters (Gaussian height and width), the reversible expansion to the ideal-gas limit must be carried out explicitly.~\cite{Ryu2008a} For IPL soft-sphere models, on the other hand, their specific scaling properties permit a formulation of the virial expansion in which a function of a single adimensional variable fully characterizes the equation of state.~\cite{Hoover1970,Young1984,Tan2010} In particular, numerical virial equations of states with up to 8 coefficients based on Monte Carlo (MC) simulations are available for exponents $n=6$, 9 and 12.~\cite{Tan2010} 

In the present paper we discuss the model proposed by Uhlenbeck and Ford~\cite{deBoer1962,Baram1991} (UF) and the scaled UF (sUF) variants as alternative reference systems for fluid-phase free-energy computations. These models are characterized by an explicitly temperature-dependent two-body interaction that, in contrast to other continuous interacting model systems, permits the exact calculation of its virial coefficients. In addition, their excess free energies with respect to the ideal gas can be represented in terms of a function of a single adimensional variable for any temperature and density, and the fluid is the only thermodynamically stable phase for a wide range of these systems. 

The remainder of the paper is organized as follows. In Sec.~\ref{Sec2} we give a brief review of the Mayer formulation for the virial equation of state and show how the specific functional form of the interactions in the UF and sUF models allows the exact calculation of the virial coefficients and leads to a particularly useful adimensional form of the equations of state. We also briefly describe the method by which the UF and sUF models can be used as reference systems in fluid-phase free-energy calculations. In Sec.~\ref{Sec3} we discuss the equations of state of the UF model, adding three more virial coefficients to the set of known values for this system~\cite{Baram1991} and provide accurate numerical representations for the equations of state and excess free energies (\verb python ~code provided in the Supplemental Material) that can be used in fluid-phase free-energy computations.  In Sec.~\ref{Sec4} we illustrate the application of the UF models as reference systems in the such calculations, considering the cases of the Lennard-Jones system and liquid silicon as described by the Stillinger-Weber model.~\cite{Stillinger1985} We conclude with a summary in Sec.~\ref{Sec5}.

\section{Theory}
\label{Sec2}
\subsection{Mayer formulation for virial equation of state}

The Mayer formulation for the virial equation of state provides a systematic approach to the computation of the virial coefficients in Eq.~(\ref{virial}). For instance, for a classical monoatomic system governed by pairwise interactions the Mayer formulation for the virial coefficients entails cluster expansions of the form~\cite{McQuarrie2000,Hansen2006}
\begin{eqnarray}
\nonumber
B_2(T)&=&-\tfrac{1}{2} \includegraphics[width=.1\columnwidth, valign=m, margin=0cm 0cm]{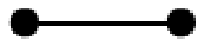}\\
\nonumber
B_3(T)&=&-\tfrac{1}{3} \includegraphics[width=.1\columnwidth, valign=m, margin=0cm 0cm]{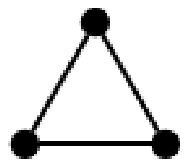} \\
\nonumber
B_4(T) &=& -\tfrac{1}{8} \left[ 3\includegraphics[width=.1\columnwidth, valign=m, margin=0cm 0cm]{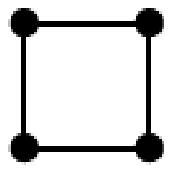} +6 \includegraphics[width=.1\columnwidth, valign=m, margin=0cm 0cm]{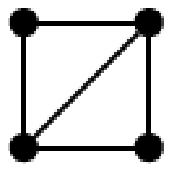} + \includegraphics[width=.1\columnwidth, valign=m, margin=0cm 0cm]{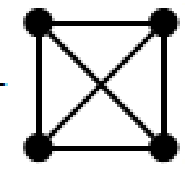} \right] \\
&\large\vdots&
\label{eq:graphexpansion}
\end{eqnarray}
in which each undirected graph represents a multidimensional integral over a specified number of particle coordinates. In particular, the virial coefficients $B_n$ involve
integrals represented by all biconnected (or star) graphs~\cite{McQuarrie2000} with $n$ vertices. In each graph, every vertex represents the coordinates of a particle and an edge connecting a pair of vertices stands for an interaction between them. For example, the graph determining $B_3$ corresponds to the integral
\begin{eqnarray}
\includegraphics[width=.15\columnwidth, valign=m, margin=0cm 0cm]{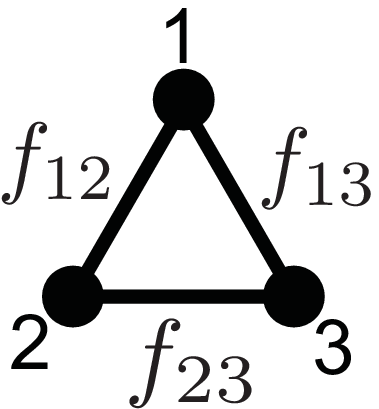} &=&  \frac{1}{V} \int_V d\textbf{r}_1\, d\textbf{r}_2 \,d\textbf{r}_3 \,f _{12} \,f _{13} \,f _{23},
\label{B3}
\end{eqnarray} 
where $\textbf{r}_i$ denotes the coordinates of particle $i$, $V$ is the system volume and $f_{ij}$ is the so-called Mayer-$f$ function~\cite{McQuarrie2000,Mayer1940} that represents the interaction between particle pair $(i,j)$. If the interactions depend only on the interparticle distance $r_{ij}$ then
\begin{equation}
f_{ij}=f(r_{ij})\equiv\exp[-\beta\, U(r_{ij})]-1,
\label{Mayer}
\end{equation}
with $U(r)$ the pair potential describing the interactions. 

In general, a biconnected graph $\mathcal{G}(n,m)$ with $n$ vertices and $m$ links then stands for the integral
\begin{equation}
I(\mathcal{G})=\frac{1}{V} \int_V d\textbf{r}_1 \cdots d\textbf{r}_n \prod_{\left<i,j\right> \in \mathcal{G}} f_{ij},
\label{I}
\end{equation} 
where the product runs over the $m$ connected pairs in $\mathcal{G}$. Although the value of the integral is independent of the particular labeling of the vertices in a graph $\mathcal{G}$, each graph appearing in the cluster expansion of Eq.~(\ref{eq:graphexpansion}) is characterized by a weight $w(\mathcal{G})=n!/|\mathcal{G}|$,~\cite{Clisby2006} with $|\mathcal{G}|$ the number of elements in the automorphism group of $\mathcal{G}$, i.e., $|\mathcal{G}|$ counts the number of distinct labelings for the topology described by $\mathcal{G}$. For instance, the three distinct biconnected graphs that contribute to $B_4$ in Eq.~(\ref{eq:graphexpansion}) are characterized by weights 3, 6 and 1, respectively. Finally, the virial coefficients $B_n$ are given by~\cite{Clisby2006}
\begin{equation}
B_n(T)=\frac{1-n}{n!}\sum_\mathcal{G} w(\mathcal{G}) I(\mathcal{G}),
\label{Bn}
\end{equation}
where the sum runs over all unlabeled biconnected graphs with $n$ vertices. 

\subsection{The Uhlenbeck-Ford model}

The difficulty in the calculation of the virial coefficients for a given model using Eq.~(\ref{Bn}) is two-fold. First, the calculations involve a potential-independent combinatorial element that amounts to counting distinct biconnected graphs that represent the possible pair-interaction topologies for a given number of interacting particles.~\cite{McQuarrie2000,Feynman1998} Second, each graph is linked to a specific many-body integral that involves products of the Mayer-$f$ function. While the diagrammatic enumerations can, in principle, be executed exactly, the associated many-body integrals in general cannot. As a result, current virial expansions for model systems such as the hard-sphere system,~\cite{Labik2005,Clisby2006,Masters2008,Wheatley2013,Schultz2014} the IPL soft-sphere model~\cite{Wheatley2013,Wheatley2005,Tan2010} and the Lennard-Jones pair potential~\cite{Singh2004,Schultz2009,Schultz2012,Ushcats2014,Feng2015} are often obtained numerically, using MC calculations to estimate the multi-dimensional integrals.

In 1962 Uhlenbeck and Ford proposed a model system featuring continuous interactions for which all the many-body integrals of the type in Eq.~(\ref{I}) can be evaluated analytically.~\cite{deBoer1962} It was named the Gaussian gas~\cite{deBoer1962,McQuarrie2000} because of the Gaussian form of its Mayer-$f$ function, 
\begin{equation}
f_{\rm UF}(r)\equiv-\exp[-(r/\sigma)^2],
\label{UF_Mayer}
\end{equation} 
where $\sigma$ is a length-scale parameter. From the definition in Eq.~(\ref{Mayer}) it follows that the associated interatomic pair potential is of the form
\begin{equation}
U_{\rm UF}(r)=-\frac{1}{\beta}\ln\left(1-e^{-(r/\sigma)^2}\right),
\label{UF_Potential}
\end{equation}
with the energy scale determined by the absolute temperature $T$. We will from now on refer to this system as the Uhlenbeck-Ford (UF) model to avoid confusion with the Gaussian-core model which is actually characterized by a Gaussian interaction potential. 

\begin{figure}[h!]
\includegraphics[width = 8.0 cm, angle=0 ]{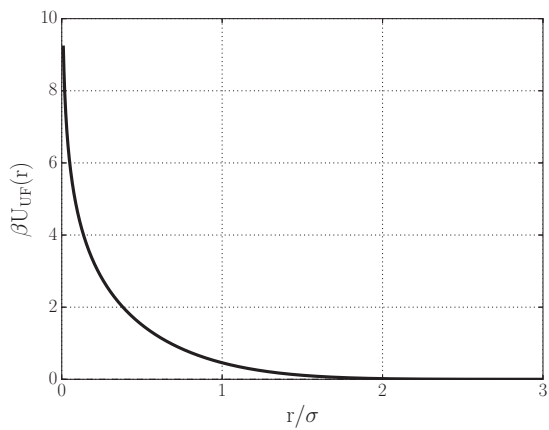}
\caption{Interatomic potential for the UF model.}
\label{Fig1}
\end{figure}

As can be seen in Fig.~\ref{Fig1} the UF model is characterized by a smooth and purely repulsive soft-sphere potential that decays rapidly for increasing distances $r$ and diverges logarithmically as $r \to 0$. 

A distinctive consequence of the particular Gaussian form of the Mayer-$f$ function in Eq.~(\ref{UF_Mayer}) is that all integrals of Eq.(\ref{I}) become multivariate Gaussian integrals that can be evaluated analytically. The results can be expressed as~\cite{deBoer1962}
\begin{equation}
I(\mathcal{G})= (-1)^m \frac{(2b)^{n-1}}{|\mathbf{M}(\mathcal{G})|^{3/2}},
\label{I_UF}
\end{equation}
with~\cite{McQuarrie2000,deBoer1962}
\begin{equation}
b\equiv\tfrac{1}{2}(\pi \sigma^2)^{3/2}
\end{equation}
and where the $(n-1)\times (n-1)$ matrix $\mathbf{M}(\mathcal{G})$ is any minor of the $n \times n$ Laplacian matrix $\mathbf{L}(\mathcal{G})$ defined as~\cite{Newman2010}
\begin{equation}
\mathbf{L}=\mathbf{D}-\mathbf{A},
\label{Laplacian}
\end{equation} 
where $\mathbf{A}(\mathcal{G})$ is the symmetric adjacency matrix~\cite{Newman2010} of the graph,
\begin{equation}
\mathbf{A}_{ij} =     \left\{ \begin{array}{rcl}
         1,  & \mbox{if vertices } i \mbox{ and } j \mbox{ are connected. } \\
         0,  & \hspace{-3.2cm}\mbox{otherwise.} \\
                \end{array}\right.
\end{equation}
and $\mathbf{D}$ is its diagonal degree matrix~\cite{Newman2010}
\begin{equation}
\mathbf{D}_{ij}=\delta_{ij}\sum_{k=1}^n \mathbf{A}_{ik},
\label{DegreeMatrix}
\end{equation}
with $\delta_{ij}$ the Kronecker delta. The absolute-value sign in the denominator of Eq.~(\ref{I_UF}) indicates the evaluation of the determinant of $\mathbf{M}$.

Combining Eqs.~(\ref{I_UF}) and (\ref{Bn}) then gives the virial coefficients 
\begin{eqnarray}
B_n &=& \frac{(1-n)\, 2^{n-1}}{n!}\, \left(\sum_\mathcal{G} w(\mathcal{G}) \, (-1)^m\,  |\textbf{M}(\mathcal{G})|^{-3/2}\right) \,b^{n-1} \nonumber \\
&\equiv&  \tilde{B}_n \,{b^{n-1}}.
\label{eq:bn_geral}
\end{eqnarray}
This result implies that the virial coefficients for the UF model are given by products of two factors, one of which ($\tilde{B}_n$) is determined exclusively by the set of biconnected graphs on $n$ vertices, whereas the other ($b^{n-1}$) encodes the length scale of the UF interaction. Also note that the $B_n$'s are independent of $T$, which is due to the temperature-dependent energy scale of the UF pair potential. 

Substitution into Eq.~(\ref{virial}) then gives the virial expansion
\begin{equation}
\beta P=\rho+\sum_{n=2}^{\infty} \tilde{B}_n b^{n-1}\rho^n,
\label{UF-virial}
\end{equation}
which can be written in adimensional form by multiplying both sides by $b$ and introducing the adimensional variable $x\equiv b\rho$, giving
\begin{equation}
\beta b P = x+ \sum_{n=2}^{\infty} \tilde{B}_n x^n.
\label{UF-virial-adim}
\end{equation}
This expression reveals that the equation of state for the UF fluid can be specified in terms of a function of a single adimensional variable $x$, regardless of the length scale $\sigma$ and temperature $T$. In addition, all coefficients $\tilde{B}_n$ can, in principle, be computed exactly. The corresponding excess Helmholtz free energy with respect to the ideal gas can be obtained by integration of Eq.~(\ref{UF-virial-adim}) to the ideal-gas limit $x=0$, giving 
\begin{equation}
\frac{\beta F_{\rm UF}^{\rm exc}(x)}{N}=\sum_{n=1}^{\infty}  \frac{\tilde{B}_{n+1}}{n} x^{n}.
\label{Helmholtz-FE-UF}
\end{equation}

\subsection{The scaled Uhlenbeck-Ford model}

Due to the logarithmic divergence of the pair potential Eq.~(\ref{UF_Potential}) for vanishing interparticle distances, the UF repulsion is extremely soft.
A convenient extension of the model that enhances the repulsion while preserving the analyticity of the multi-center integrals in the virial expansion is obtained by scaling the energy unit $k_B T$  by an integer $p>1$. This defines the scaled Uhlenbeck-Ford (sUF) potential-energy function
\begin{equation}
U^{(p)}_{\rm UF}(r)=-\frac{p}{\beta}\ln\left(1-e^{-(r/\sigma)^2}\right),
\label{Scaled_UF_Potential}
\end{equation}
and the corresponding Mayer-$f$ function 
\begin{eqnarray}
\nonumber
f^{(p)}_{\rm UF}(r)&=&\exp\left[p \ln\left(1-e^{-(r/\sigma)^2}\right)\right]-1 \\
\nonumber
 &=& \left(1-e^{-(r/\sigma)^2}\right)^p-1 \\
\label{Scaled_UF_Mayer}
&=& \sum_{k=1}^p (-1)^{k}\, \binom pk \, e^{-k (r/\sigma)^2}.
\end{eqnarray}

The integral $I^{(p)}(\mathcal{G})$ associated with each biconnected graph $\mathcal{G}$ with $n$ vertices and $m$ edges then becomes
\begin{eqnarray}
\nonumber
I^{(p)}(\mathcal{G})&=&\frac{1}{V} \int_V d\textbf{r}_1 \cdots d\textbf{r}_n \prod_{\left<i,j\right> \in \mathcal{G}} f^{(p)}_{ij} \\
\nonumber
&=& \frac{1}{V} \int_V d\textbf{r}_1 \cdots d\textbf{r}_n \sum_{k_1=1}^p \cdots \sum_{k_m=1}^p (-1)^{k_1+\, \cdots\, +k_m} \times \\
&\ & \binom {p}{k_1} \cdots \binom {p}{k_m} \exp\left[-\sum_{l=1}^m k_{l} r_l^2/\sigma^2\right],
\label{Iscaled}
\end{eqnarray} 
where the index $l \in [1,m]$ identifies the $m$ edges of the graph and $r_l$ denotes the distance between the interacting particle pair ($i,j$) associated with a given edge $l$. 

The integral in Eq.~(\ref{Iscaled}) is thus a sum of $p^m$ multivariate Gaussian integrals, each of which is given by an expression similar to that in Eq.~(\ref{I_UF}). Specifically, we have
\begin{eqnarray}
\nonumber
I^{(p)}(\mathcal{G})&=&\sum_{k_1=1}^p \cdots \sum_{k_m=1}^p (-1)^{k_1+\, \cdots\, +k_m} \ \binom {p}{k_1} \cdots \binom {p}{k_m} \times \\ 
&\ &  \frac{(2b)^{n-1}}{|\mathbf{M}(\mathcal{G},\{k_1, \dots, k_m \})|^{3/2}},
\end{eqnarray} 
where the matrix $\mathbf{M}(\mathcal{G},\{k_1, \dots, k_m \})$ is any minor of the Laplacian matrix $\mathbf{L}(\mathcal{G},\{k_1, \dots, k_m \})$ associated with the \emph{weighted} graph obtained by attributing weights $k_l$ to the $m$ edges of $\mathcal{G}$.  $\mathbf{L}(\mathcal{G},\{k_1, \dots, k_m \})$ continues to be defined according to Eq.~(\ref{Laplacian}) but now with $\mathbf{A}(\mathcal{G},\{k_1, \dots, k_m \})$ being the \emph{weighted} adjacency matrix,
\begin{equation}
\mathbf{A}_{ij} =     \left\{ \begin{array}{rcl}
         k_l,  & \mbox{if vertices } i \mbox{ and } j \mbox{ are connected by edge } l. \\
         0,  & \hspace{-4.6cm}\mbox{otherwise.} \\
                \end{array}\right.
\end{equation}
The degree matrix $\mathbf{D}$ continues to be defined as in Eq.~(\ref{DegreeMatrix}). Evidently, for $p=1$ all weights become equal to unity and the above result reduces to Eq.~(\ref{I_UF}).

Combination of Eqs.~(\ref{Bn}) and (\ref{Iscaled}) then gives the virial coefficients $B^{(p)}_n$ for the scaled UF model,
\begin{eqnarray}
\nonumber
B^{(p)}_n &=& \frac{(1-n)2^{n-1}}{n!}\, 
\sum_\mathcal{G} w(\mathcal{G}) \left( \sum_{k_1=1}^p \cdots \sum_{k_m=1}^p (-1)^{k_1+\, \cdots\, +k_m} \times \right. \\ 
\nonumber
&\ & \left. \binom {p}{k_1} \cdots \binom {p}{k_m} \ 
\frac{1}{|\mathbf{M}(\mathcal{G},\{k_1, \dots, k_m \})|^{3/2}} \right) b^{n-1}\\
&\equiv&  \tilde{B}^{(p)}_n \,{b^{n-1}}.
\label{eq:bn_scaled}
\end{eqnarray}
The general structure of the $B^{(p)}_n$'s remains the same as that for the UF model, with the virial coefficients being products of a factor that depends only on the properties of the set of biconnected graphs and another that depends only on the length scale $\sigma$ of the interaction. As a result, the virial equation of state and excess Helmholtz free-energy expressions for the sUF model preserve all properties of the UF model and are analogous to those in Eqs.~(\ref{UF-virial}) and (\ref{Helmholtz-FE-UF}),   
\begin{equation}
\beta b P = x+ \sum_{n=2}^{\infty} \tilde{B}^{(p)}_n x^n,
\label{UF-virial-adim_scaled}
\end{equation}
and
\begin{eqnarray}
\frac{\beta F^{\rm exc (p)}_{\rm UF}(x)}{N} &=& \sum_{n=1}^{\infty}  \frac{\tilde{B}^{(p)}_{n+1}}{n} x^{n}.
\label{Helmholtz-FE-scaled_UF}
\end{eqnarray}

\subsection{UF and sUF models as reference systems in free-energy calculations}
\label{FreeEnergy}

Because the excess Helmholtz free energies of the UF and sUF fluids can be determined exactly, in principle, these models are promising candidates to serve as reference systems in fluid-phase free-energy computations. Such calculations generally determine the free-energy difference between a system of interest and a reference by estimating the work $W_{\rm rev}$ associated with a reversible process between the two. This is often accomplished by defining a parametrized Hamiltonian
\begin{equation}
\label{interpolation}
H(\lambda)=(1-\lambda) H_{\rm ref} + \lambda H_{\rm int},
\end{equation} 
with $H_{\rm int}$ and $H_{\rm ref}$ the Hamiltonians of the systems of interest and reference, respectively. The free-energy difference between them is then obtained by the integral
\begin{equation}
\Delta F\equiv F_{\rm int}-F_{\rm ref}=W_{\rm rev}=\int_0^1\left\langle \frac{\partial H}{\partial \lambda}\right\rangle_{\lambda} d\lambda,
\label{TI}
\end{equation}
which represents the reversible work done by the generalized driving force $\partial H/\partial \lambda$ associated with the parameter $\lambda$, and where the brackets indicate an equilibrium ensemble average. Provided the free energy of the reference system is known, the absolute free energy of the system of interest is then given by
\begin{equation}
F_{\rm int}=F_{\rm ref}+W_{\rm rev}.
\end{equation}  
In this context, the UF and sUF models appear ideally suited to serve as such reference systems for fluid-phase systems and this will be further investigated in the remainder of the paper. 

\section{Equations of state and excess Helmholtz free energies}
\label{Sec3}
\subsection{UF model}
%\subsubsection{Virial Equation of State}

To obtain an excess Helmholtz free energy expression that is as accurate as possible, we determine the virial coefficients $\tilde{B}_n$ with $n$ as large as achievable. To this end, we evaluate expression Eq.~(\ref{eq:bn_geral}) by explicitly generating and determining the weights for all unlabeled biconnected graphs $\mathcal{G}$ for given $n$ using the \verb nauty ~package.~\cite{McKay2014} The integer-element matrix determinants are computed using the Bareiss algorithm,~\cite{Bareiss1968} the execution time of which scales as $O(n^3)$. This procedure allows us to compute the exact virial coefficients up to $\tilde{B}_{13}$, as summarized in Table~\ref{tab:1}. 
\begin{table}[h!tb]
\caption{Numerical representation of the exact virial coefficients for the UF model up to $\tilde{B}_{13}$.}
\centering
\begin{tabular}{r r r} 
\\
\hline
\hline \\
\vspace{0.1cm}
$n$ &  Graphs  & $\tilde{B}_n$ \\
\hline 
\\
    2   &  1 & 1 \\ \vspace{0.02cm} 
    3    &  1 &           0.25660011963983367311 \\ \vspace{0.02cm}
    4    &  3 &          -0.12545995705504467834 \\ \vspace{0.02cm}
    5   &  10 &           0.01332565517320544100 \\ \vspace{0.02cm}
    6    &  56 &          0.03846093583086967155 \\ \vspace{0.02cm}
    7    &  468 &        -0.03308344290314971739 \\ \vspace{0.02cm}
    8    &  7123 &        0.00418241876968238735 \\ \vspace{0.02cm}
    9    &  194066 &      0.01519760719550087466 \\ \vspace{0.02cm}
    10    &  9743542 &   -0.01384965413457514293 \\ \vspace{0.02cm}
    11    & 900969091  &  0.00133475791711096611 \\ \vspace{0.02cm}
    12    & 153620333545  &  0.00760432724881212587 \\ \vspace{0.02cm}
    13    & 48432939150704 & -0.00672644140878158825 \\
\hline
\end{tabular}
\label{tab:1}
\end{table}
The values of $\tilde{B}_{10}$ and $\tilde{B}_{13}$, to the best of our knowledge, are reported here for the first time and they continue the previously observed pattern displaying two positive coefficients followed by a negative one.~\cite{Baram1991} 

To provide a sense of the computational effort involved in these calculations, the evaluation of $\tilde{B}_{13}$, which requires the generation and processing of $\sim~10^{13}$ graphs, required more than 11 days of walltime using 780 CPU cores. Computation of the next coefficient, $\tilde{B}_{14}$, would take $\sim 600$ times longer and is currently out of reach.

\begin{figure} [h!]
\includegraphics[width = 8.0 cm, angle=0 ]{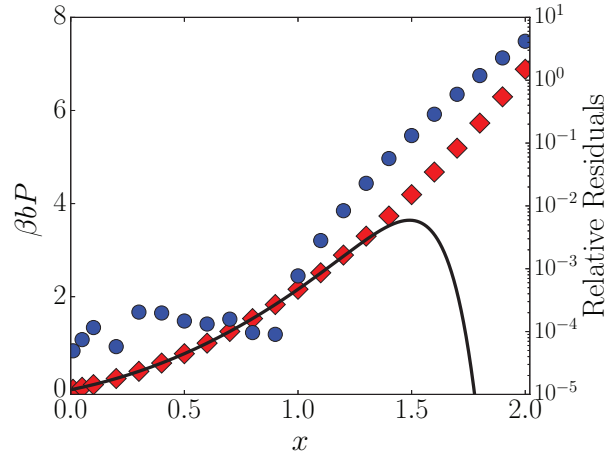}
\caption{Comparison of the scaled pressure as a function of the adimensional variable $x$ for the UF model as obtained from the exact virial expansion up to $\tilde{B}_{13}$ using the values in Table~\ref{tab:1} (full line) and MD simulations (red diamonds), respectively. Relative residuals are depicted by blue circles. The statistical uncertainty in the MD averages is smaller than the symbol size for all values of $x$.}
\label{Fig2}
\end{figure}

To assess the accuracy of expansion Eq.~(\ref{UF-virial-adim}) based on the first 13 virial coefficients, we compare it to values obtained from constant NVT MD simulations using the \verb LAMMPS ~code.~\cite{Plimpton1995} We choose the UF parameters to be $\sigma=\sqrt{(2)^{2/3}/\pi}$ and $\beta=1$, and employ a cut-off radius of $r_c=5 \sigma$, without shifting.
Using a cubic periodic simulation cell containing 864 particles and integrating the equations of motion for the Langevin thermostat for $2\times 10^7$ time steps with $\Delta t=0.003$ in reduced units and a damping time scale of 100$\Delta t$, we compute the mean scaled pressure as a function of the adimensional variable $x$ by carrying out a series of simulations at different number densities $\rho$. The results, displayed in Fig.~\ref{Fig2}, are in good agreement with the truncated virial expansion for values of $x\lesssim 1$, with relative errors smaller than $1\times10^{-3}$. For larger values, on the other hand, it quickly diverges from the MD data. Two factors that hamper the rate of convergence of the virial expansion are the slow decay of the absolute values of the $\tilde{B}_n$, with $\tilde{B}_{13}$ still being of the same order of magnitude as $\tilde{B}_8$, and the alternating character of their signs.

\begin{figure} [h!]
\includegraphics[width = 8.0 cm, angle=0 ]{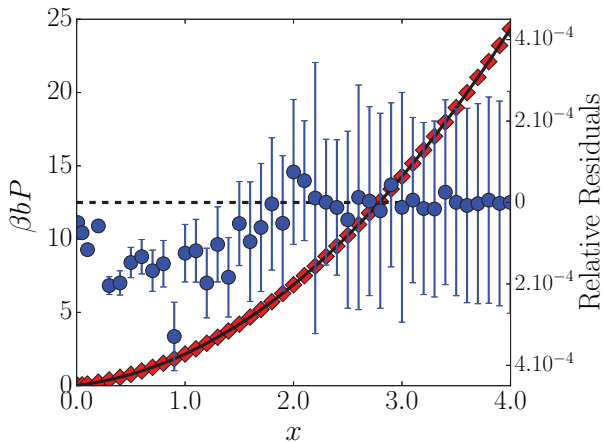}
\caption{Comparison of the scaled pressure as a function of the adimensional variable $x$ for the UF model as obtained from the fitted virial expansion up to $\tilde{B}_{22}$, using the values in Table~\ref{tab:1} for coefficients $\tilde{B}_1$ to $\tilde{B}_{13}$ (black line), and MD simulations (red diamonds). Blue circles depict relative residuals between MD data and fitted virial expansion. Error bars represent relative standard error in the MD averages.}
\label{Fig3}
\end{figure}

One possibility to extend the representation of the MD pressure data in terms of the virial equation of state for $x>1$ is to estimate a larger number of virial coefficients, fixing the first 13 coefficients to their exact values and fitting higher-order $\tilde{B}_n$'s in a least-squares procedure. Results of such a procedure are shown in Fig.~\ref{Fig3}, which compares MD data for $x\leq 4$ to a fitted virial expansion up to order 22. 
Overall, the agreement between MD data and the fitted virial expansion is very good, in particular for the larger values of $x$, where the relative residuals fluctuate around zero and are of the order of $10^{-7}$. Still, for $x \lesssim 1.5$ the deviations between the fitted truncated virial expansion and the MD results are significantly larger and systematically negative. In addition to the above-mentioned convergence issues of the truncated virial expansion, these discrepancies may also be related to the global nature of the regression procedure in which the total sum of the squared deviations across the entire interval of $x$-values is minimized. In any event, these results suggest that the truncated virial expansion may not be the best representation for the equation of state of the UF model, in particular when keeping in mind the objective of using this model as a reference system in fluid-phase free-energy calculations, as detailed in Sec.~\ref{FreeEnergy}.

We improve the numerical representation of the MD data by adopting a piece-wise fitting procedure of the equation of state using cubic splines.~\cite{Press2007} In this fashion the fit passes through all MD data points and provides a smooth interpolation scheme for the intermediate values of $x$. Moreover, as discussed further below, this representation also leads to an excellent description for the excess Helmholtz free energy function $F^{\rm exc}_{\rm UF}(x)$ which will be used in fluid free-energy calculations as discussed in Sec.~\ref{FreeEnergy}. All spline representations of equations of state and excess Helmholtz free energies described in the paper are available by means of the \verb python ~script \verb ufGenerator.py ~supplied in the Supplemental Material.

\begin{figure} [h!]
\includegraphics[width = 8.0 cm, angle=0 ]{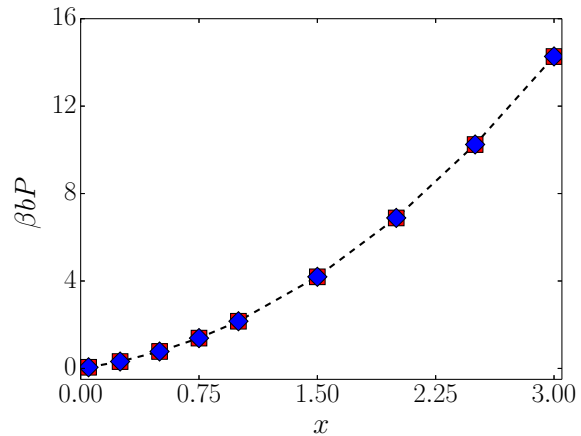}
\caption{Comparison of the scaled pressures as a function of the adimensional variable $x$ for the UF model using two different values of the length-scale parameter $\sigma$, with the corresponding $b$ value of one (circles) being twice as large as that of the other (diamonds). Error bars are smaller than symbol size. Dashed line serves as a guide to the eye.}
\label{Fig4}
\end{figure}

As a further consistency check, we also verify numerically the scaling property encoded in Eq.~(\ref{UF-virial-adim_scaled}). For this purpose, we plot the adimensional quantity $\beta b P$ as a function of $x$ for the values $b=1$ and 2, respectively.
The resulting comparison, obtained using MD simulations, is displayed in Fig.~\ref{Fig4} and it confirms the scaling relationship.  

\begin{figure} [h!]
\includegraphics[width = 8.0 cm, angle=0 ]{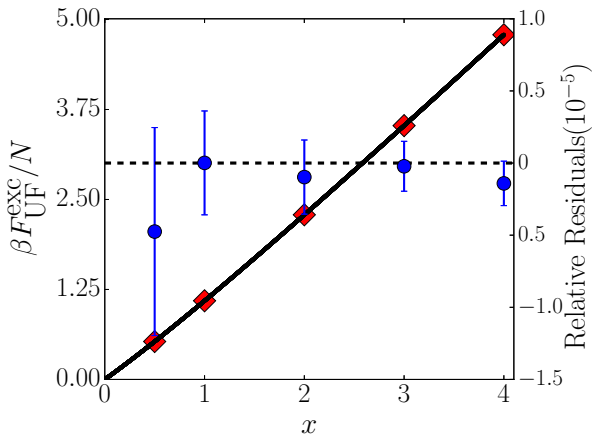}
\caption{$F^{\rm exc}_{\rm UF}(x)$ as obtained by integrating the cubic spline representation of the equation of state (full line). Red diamonds points depict results for $F^{\rm exc}_{\rm UF}(x)$ obtained from switching free-energy calculations (see text). Blue circles denote relative deviation of integrated cubic-spline excess free-energy compared to results of switching free-energy calculations. Dashed line depicts level of zero residual.}
\label{Fig5}
\end{figure}

The excess Helmholtz free energy of the UF fluid with respect to the ideal gas as a function of $x$ can be obtained by standard thermodynamic integration of the pressure with respect to the volume.~\cite{Frenkel2002} We have performed this integration using the cubic-spline representation of the equation of state obtained in the previous Section, and the result is shown in Fig.~\ref{Fig5} as the full line. To assess the accuracy of this approach we also compute $F^{\rm exc}_{\rm UF}$ explicitly using the method described in Sec.~\ref{FreeEnergy}. To this end, we take the thermodynamic path of Eq.~(\ref{interpolation}) to be a linear interpolation between the UF model and the ideal gas: 
\begin{equation}
{H(\lambda)}=\lambda H_{\rm UF}+(1-\lambda) H_{\rm i.g.}=K+\lambda U_{\rm UF}, 
\label{path}
\end{equation}
with $K$ the kinetic energy and $\lambda$ varying between 0 and 1. Notice that, in contrast to many other pair interactions, the driving force $\partial H/\partial \lambda = U_{\rm UF}$ on this path is finite everywhere. In particular, in the ideal gas limit described by $H(0)$, its mean value given by~\cite{Hansen2006}
\begin{eqnarray}
\nonumber
\left\langle U_{\rm UF}\right\rangle&=&2 \pi N \rho\int_0^{\infty}r^2 g(r) U_{\rm UF}(r) \,dr \\
\nonumber
&=& -\frac{2 \pi N \rho}{\beta}\int_0^{\infty}r^2 \ln\left(1-e^{-(r/\sigma)^2}\right) \,dr \\
&=& \frac{N \,x}{\beta} \,\zeta\left(\frac{5}{2}\right),
\end{eqnarray}
where $N$ is the number of particles, $g(r)=1$ is the radial distribution function for the ideal gas, $x$ is the same adimensional variable used in Eq.~(\ref{UF-virial-adim}) and $\zeta(s)$ is the Riemann zeta function. 

To compute the reversible work $W_{\rm rev}$ we employ the adiabatic switching (AS) approach,~\cite{Watanabe1990,deKoning1997,Freitas2016} in which the integral over equilibrium ensemble averages, Eq.(\ref{TI}), is replaced by a time integral over a dynamic, intrinsically non-equilibrium process in which the interpolation parameter $\lambda=\lambda(t)$ is time dependent and the integration is performed over instantaneous values of the driving force. The associated dynamical work is given by
\begin{equation}
W_{\rm dyn}=\int_0^{t_s}\frac{d\lambda}{dt} \left.\frac{\partial H}{\partial \lambda}\right|_{\lambda(t)}  dt,
\label{W_dyn}
\end{equation}
in which the integration involves instantaneous values of $\partial H/\partial \lambda$ rather than the ensemble averages in Eq.~(\ref{TI}) and $t_s$ is the switching time.
 
In this manner, AS provides an estimate of the reversible-work integral from a single non-equilibrium, instead of a series of independent equilibrium simulations. The price to pay for this reduction of computational effort is that $W_{\rm dyn}$ is a stochastic variable whose mean value, by the second law of thermodynamics, is shifted with respect to $W_{\rm rev}$ according to 
\begin{equation}
\overline{W_{\rm dyn}}\geq W_{\rm rev},
\label{Wdyn}
\end{equation}
where the overbar denotes an average over different realizations of the nonequilibrium process and the equality is valid only in the reversible limit where the dynamic process is executed infinitely slowly. Nevertheless, when the process is sufficiently slow for linear-response theory to be accurate, it can be shown~\cite{deKoning1997} that the systematic error implicit in Eq.~(\ref{Wdyn}) can be eliminated by combining the results of the processes realized in both directions: $H_{\rm ref} \to H_{\rm sys}$ and $H_{\rm sys} \to H_{\rm ref}$,~\cite{deKoning2005,Freitas2016}
giving the unbiased estimator
\begin{equation}
\Delta F=\tfrac{1}{2} \left[ \overline{W_{\rm dyn}({\rm ref} \to {\rm sys})}-\overline{W_{\rm dyn}({\rm sys} \to {\rm ref})}\right].
\label{workaverage}
\end{equation}

Applying this scheme, we use 10 independent process realizations for each direction to estimate the averages and uncertainties in Eq.~(\ref{workaverage}). The specifications of the corresponding MD simulations are identical to those used to obtain the results in Figs.~\ref{Fig2} and \ref{Fig3}. The non-equilibrium processes are characterized by a linear switching function $\lambda(t)$ and the initial state of each switching process is first equilibrated for $t_{\rm eq}=10^5 \Delta t$. The switching time was chosen to be $t_s=5\times10^6 \,\Delta t$, which is sufficiently long to produce non-equilibrium processes in the linear-response regime. 

The free-energy results are depicted in Fig.~\ref{Fig5}, in which the full line represents the excess Helmholtz free energy obtained from the integration of the cubic-spline equation of state and the data points denote values obtained from the explicit free-energy calculations.
The agreement is excellent throughout, with the cubic-spline-based results identical to the explicit free-energy data within the error bars. 

As a final comment we turn to the issue of phase stability of the UF model. Indeed, previous mathematical analysis~\cite{Baram1990,Baram1991} of the convergence properties of the virial equation of state for the UF model has shown that, due to the extreme softness of the interactions, the fluid is most likely the only stable phase of the system, regardless of interaction length scale and density. In other words, the UF system does not undergo a fluid-solid transition for any choice of parameters and this is evidently a favorable characteristic for a fluid-phase reference system. Indeed, it provides greater flexibility as a fluid-phase reference system compared to, for example, the IPL model. Although the equation of state of the latter can also be expressed in an adimensional form, its phase diagram features a prominent presence of crystalline phases.~\cite{Prestipino2005} 

On the other hand, even though the fluid is the only stable phase of the UF model, when attempting to compute the free energy of a liquid phase close to a solid-liquid-vapor triple point, for instance, one cannot \emph{a priori} guarantee that all intermediate Hamiltonians in Eq.~(\ref{interpolation}) remain outside their respective liquid-vapor coexistence regions. If not, the switching process may possibly be hampered by liquid-vapor coexistence along the way. Nonetheless, this is a general issue for switching processes described by Hamiltonians of the type in Eq.~(\ref{interpolation}) and is not specific to the case in which the UF model is used as a reference system.

\subsection{Scaled UF model}

Due to its extreme softness, the applicability of the UF model as a reference system for fluid-phase free-energy calculations is limited, as will become clear momentarily. However, as discussed in the previous section, it is possible to increase the repulsion of the UF model while preserving the analyticity of its virial coefficients and the adimensional form of the corresponding equation of state by multiplying the potential-energy expression by a positive integer $p>1$. The virial coefficients are then given by Eq.~(\ref{eq:bn_scaled}) and they can be computed using a procedure similar to that followed to compute the virial coefficients in Table~\ref{tab:1}. However, the computational overhead associated with this procedure quickly becomes prohibitively large compared to the case $p=1$. This is due to the fact that, for $p>1$, each of the biconnected graphs $\mathcal{G}$ with $m$ edges contributes with $p^m$ terms in Eq.~(\ref{eq:bn_scaled}). For example, the complete graph on $n$ vertices has $m=n(n-1)/2$ edges, giving $p^m=p^{(n^2-n)/2}$ terms. For $n=5$ and $p=3$, for instance, this entails the calculation of $3^{10}=9,765,625$ contributions, just for this single graph. Given the exponential growth of the number of distinct biconnected graphs with increasing $n$, as shown in Table~\ref{tab:1}, it is clear that the exact calculation of the $\tilde{B}^{(p)}_n$'s rapidly becomes impracticable. This is reflected in Table~\ref{tab:3}, which shows numerical representations of the exact values of $\tilde{B}^{(p)}_n$ for the particular cases of $p=2$ and $p=10$. For $p=10$, the evaluation of $\tilde{B}^{(10)}_6$ is already beyond reach, requiring several months of processing time on hundreds of CPUs.
\begin{table}[h!tb]
\caption{Exact virial coefficients for the scaled UF model for $p=2$ and $p=10$.}
\centering
\vspace{0.1cm}
\begin{tabular}{r r r}
\hline
\hline
\vspace{0.1cm}
$n$ & $\tilde{B}^{(2)}_n$   & $\tilde{B}^{(10)}_n$ \\
\hline 
    2   &  1.64644660940 &   4.01438397540\\ \vspace{0.02cm} 
    3    & 0.94319582715 &   8.03162517036\\ \vspace{0.02cm}
    4    &-0.33498518088 &   5.79111386879 \\ \vspace{0.02cm}
    5   & -0.30368857400 &  -7.65837103727 \\ \vspace{0.02cm}
    6    & 0.39585778168 &    \\ \vspace{0.02cm}
    7    & 0.05365513098 &    \\ \vspace{0.02cm}
    8    &-0.40848140600 &    \\
\hline
\end{tabular}
\label{tab:3}
\end{table}

Even though the exact calculation of virial coefficients for the scaled UF model rapidly becomes unfeasible, the fact that its equation of state Eq.(\ref{UF-virial-adim_scaled}) preserves the adimensional form of the original UF model enables one to obtain numerically accurate equations of state by fitting MD data of the reduced pressure $\beta b P$ as a function of the single variable $x$ using the same cubic-spline approach adopted above. 

\begin{figure} [h!]
\includegraphics[width = 8.0 cm, angle=0 ]{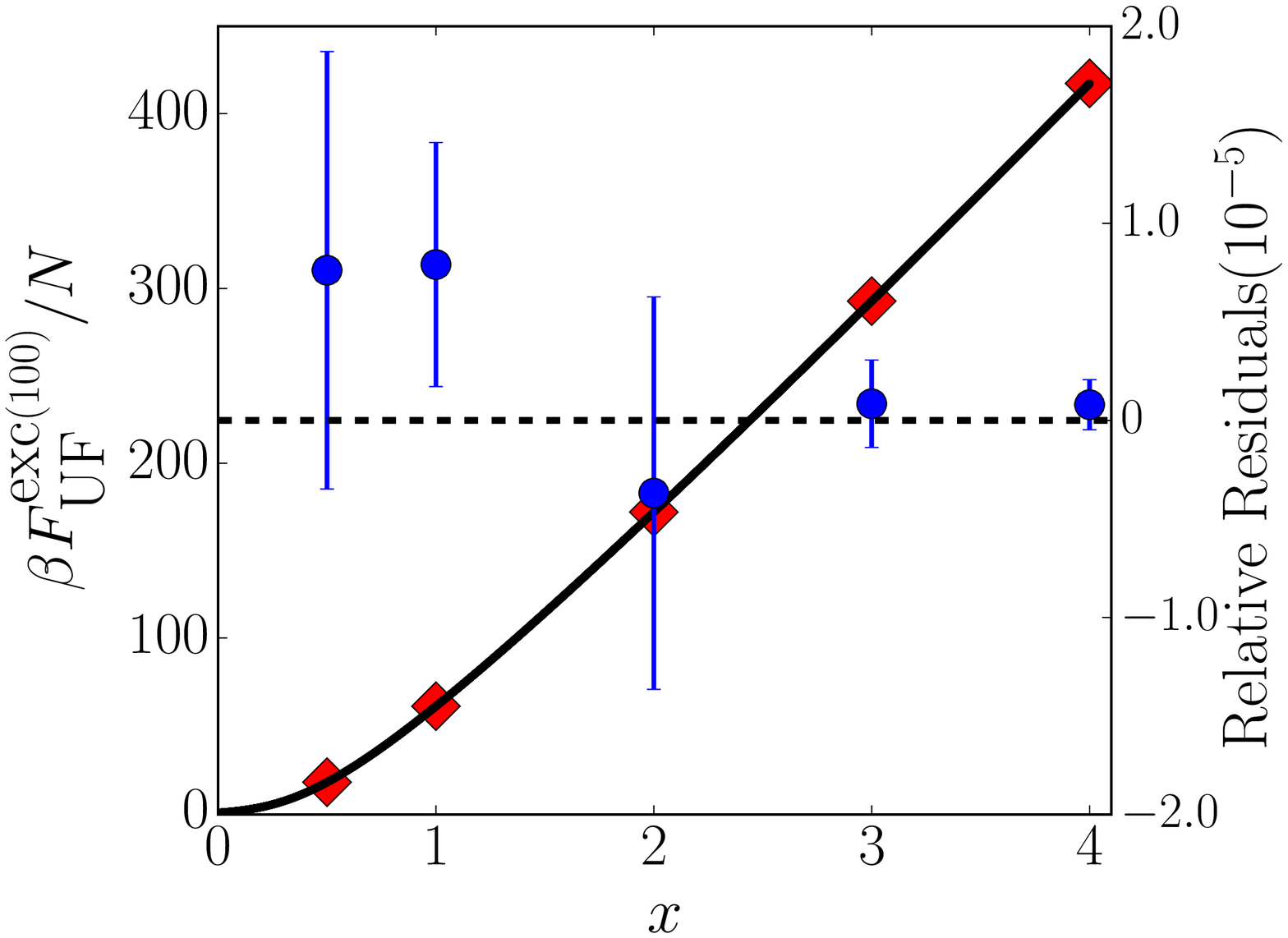}
\caption{$F^{\rm exc (100)}_{\rm UF}(x)$ as obtained by integrating the cubic spline representation of the equation of state (full line). Red diamonds points depict results for $F^{\rm exc (100)}_{\rm UF}(x)$ obtained from the switching free-energy calculations. Blue circles denote relative deviation of integrated cubic-spline excess free-energy compared to switching free-energy calculations. Dashed line depicts level of zero residual.}
\label{Fig6}
\end{figure}

In addition, using these equation-of-state cubic-spline representations, we compute the excess Helmholtz free energy for these scaled UF models through direct integration of the thermodynamic relations. Figure~\ref{Fig6} displays the resulting $F^{\rm exc (p)}_{\rm UF}(x)$ for $p=100$ and compares it to results of switching free-energy calculations using the thermodynamic path given by Eq.(\ref{path}). The results are again very good, with the integrated equation-of-state values in excellent  agreement with the switching results throughout the entire interval of $x$-values. All cubic-spline based results for the equations of state as well as the excess Helmholtz free energy expressions for $p=25$, 50, 75 and 100 have been implemented in the \verb python ~script \verb ufGenerator.py ~supplied in the Supplemental Material.

A remaining concern involves the issue of fluid-phase stability for the sUF models. As mentioned in the previous subsection, the UF model most likely exists only in the fluid phase, irrespective of thermodynamic conditions and model parameters. For the sUF models, on the other hand, it is to be expected that beyond some minimum value of the scaling factor $p$ crystalline phases become stable under particular thermodynamic conditions due to the reduced softness of the repulsion. The information regarding such conditions is contained in the phase diagram of the sUF model in terms of the scaling factor $p$ and the variable $x$. Calculations to determine this phase diagram are currently in progress and the results will be reported elsewhere. However, preliminary data indicate that for $p \lesssim 100$ the fluid remains the only stable phase of the sUF model for any $x$. 

\begin{figure} [h!]
\includegraphics[width = 9.0 cm, angle=0 ]{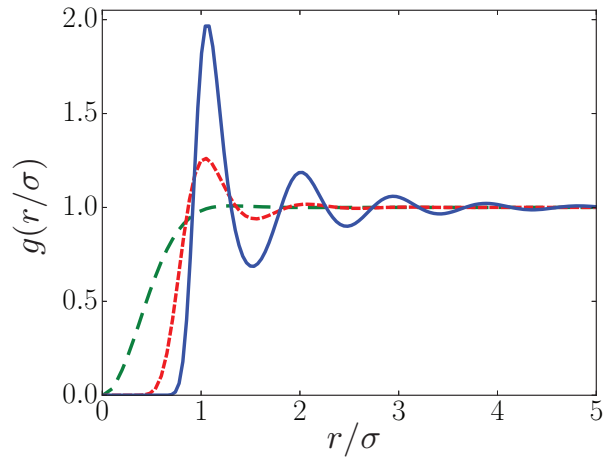}
\caption{Radial distribution functions for the sUF models with $p=1$ (long-dashed line), $p=10$ (short-dashed line) and $p=50$ (full line), respectively.}
\label{Fig7}
\end{figure}

Finally, it is interesting to assess the effect of the scaling factor $p$ on the structure of the UF and sUF fluids. To this end we compute the radial distribution functions for the sUF models characterized by $p=1$, $p=10$ and $p=50$, respectively, for $x=1$ and the results are displayed in Fig.~\ref{Fig7}. It is clear that the UF fluid essentially has no structure at all, with its radial distribution function monotonically rising to the ideal-gas plateau without displaying any peaks. The effect of increasing the repulsion by introducing the scaling factors is a progressive enhancement of liquid structure, with the fluid for $p=10$ showing a clear first solvation shell while for the case of $p=50$ the fluid displays 4 coordination shells.

\section{Applications}
\label{Sec4}
To assess their applicability in fluid-phase free-energy calculations, we compute the free energies of two different fluids using UF models as reference systems and compare the results to those obtained using independent calculations. Specifically, we compute the excess Helmholtz free energies of a Lennard-Jones fluid, which is characterized by adirectional two-body interactions, and a liquid state of the Stillinger-Weber model for Si,~\cite{Stillinger1985} which involves directional three-body contributions. All the free-energy calculations are carried out using MD simulations as implemented in the \verb LAMMPS ~package.~\cite{Plimpton1995}

\subsection{Lennard-Jones fluid}

For the case of the LJ pair potential $U_{\rm LJ}$, with length and energy scales $\sigma$ and $\epsilon$, respectively, we compute the excess Helmholtz free energy for the fluid state characterized by an inverse temperature of $\beta=0.5 \epsilon^{-1}$ and a density of$\rho=0.1\,\sigma^{-3}$. The LJ interaction potential was truncated and shifted at $r_c=4.0 \sigma$.

The excess free energy $\Delta F$ was determined using two different thermodynamic paths. In the first, we employ the UF model as a reference system, using the parametrized Hamiltonian
\begin{eqnarray}
\nonumber
H_1(\lambda)&=&\lambda H_{\rm LJ}+(1-\lambda) H_{\rm UF} \\
&=&K+\lambda U_{\rm LJ}+(1-\lambda) U^{(p)}_{\rm UF},
\label{H1}
\end{eqnarray}  
with $K$ the kinetic energy.
Accordingly, the excess free energy of the LJ system can be computed by
\begin{eqnarray}
\nonumber
&\Delta& F\equiv F_{\rm LJ}-F_{\rm i.g.} \\
&=& W_{\rm rev}(\rm UF\to LJ) + F^{\rm exc (p)}_{\rm UF},
\label{Wrev1}
\end{eqnarray}
and $F^{\rm exc (p)}_{\rm UF}$ is the excess Helmholtz free energy of the sUF model as given by the integrated cubic-spline representation of the UF equations of state described in the previous Section. 

The second thermodynamic path, as discussed in Sec.~\ref{Sec1}, is composed of two steps, making use of an intermediate reference between the LJ system and the ideal gas. Here we choose the Gaussian core (GC) pair potential~\cite{Stillinger1976,Prestipino2005,Prestipino2005a,Ryu2008a}, given by
\begin{equation}
U_{\rm GC}(r)=\epsilon_{\rm GC}\exp\left[-\left(\frac{r}{\sigma_{\rm GC}}\right)^2\right],
\end{equation}
for this purpose. In this fashion, we define the parametrized Hamiltonians
\begin{equation}
{H_{2a}=\lambda H_{\rm LJ}+(1-\lambda) H_{\rm GC}}
\label{H2a}
\end{equation}
and 
\begin{equation}
{H_{2b}=\lambda H_{\rm GC}+(1-\lambda) H_{\rm i.g.}}=K+\lambda U_{\rm GC}.
\label{H2b}
\end{equation}
Determining the reversible work associated with both paths using the AS approach above, the excess Helmholtz free energy is computed as
\begin{equation}
\Delta F=W_{\rm rev}(\rm GC\to LJ)+W_{\rm rev}(\rm i.g.\to GC)
\label{Wrev2}
\end{equation}
The explicit calculation of the excess Helmholtz free energy of the GC system is necessary because, in contrast to the case of the UF model (as shown in Fig.~\ref{Fig4}), the equations of state for different values of the model parameters cannot be rescaled to collapse onto a single master curve.

We use the AS approach to compute the reversible work values in Eqs.~(\ref{Wrev1}) and (\ref{Wrev2}), and the parameters of the associated MD simulations are identical to those employed in Sec.~\ref{Sec3}. 
\begin{figure} [h!]
\includegraphics[width = 8.0 cm, angle=0 ]{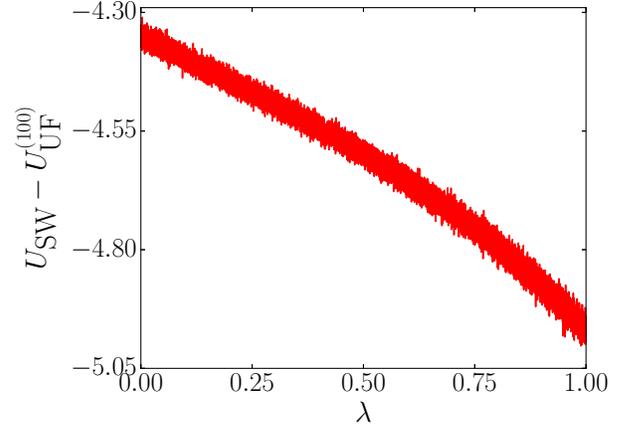}
\caption{Driving force $U_{\rm LJ}-U^{(p)}_{\rm UF}$ as a function of $\lambda$ for $p=1$ (A) and $p=50$ (B), with $\sigma_{\rm UF}=0.58 \sigma$ and $\beta=0.5 \epsilon^{-1}$. Inset focuses on divergent driving force close to $\lambda=0$ for the case $p=1$.}
\label{Fig8}
\end{figure}
Focusing on the first thermodynamic path described by the Hamiltonian in Eq.~(\ref{H1}), we initially use the original UF model characterized by $p=1$, with $\sigma_{\rm UF}=0.58 \sigma$ (which corresponds to $x=0.05432$ at the considered density), $\beta=0.5$ and a cutoff distance of $r_c=4 \sigma_{\rm UF}$. Fig.~\ref{Fig8}~A shows the evolution of the corresponding driving force along a typical switching simulation for this case. There is manifest divergent behavior close to $\lambda=0$, which is a consequence of the extreme softness of the UF model. As a result, for values near $\lambda=0$ when the dynamics is predominantly governed by the UF model, the interparticle distances can become very small, leading to very large positive values of the driving force because of the $U_{\rm LJ}$ contribution. For this reason, the original UF model is not adequate as a reference system in fluid-phase free-energy calculations. This divergence problem, however, can be resolved by using the sUF model. This is demonstrated in Fig.~\ref{Fig8}~B, which depicts the driving force for the same process, but now using the sUF model with $p=50$. The results clearly demonstrate the versatility of the sUF model. The average driving force is now a smooth function of $\lambda$, with the fluctuations remaining small throughout the entire range of $\lambda$-values. Moreover, its magnitude is much smaller than that of the UF model and it has the same order of magnitude along the entire process. Thus, the adjustable energy scale of the sUF model allows one to obtain a much better reference system compared to the original UF model, but still preserving all analytical properties of the UF virial equation of state.   

For the sUF system (with $p=50$) as the reference system, the excess Helmholtz free energy for the LJ liquid at the specified conditions is found to be
\begin{equation*}
\frac{\beta F^{\rm exc}_{\rm LJ}}{N}=(-0.10824 \pm 0.00005),
\end{equation*}
where the error bar represents one standard deviation in the mean for the unbiased dynamical estimator as obtained from 10 independent realizations of the AS process in both directions

Considering the second thermodynamic path described by the switching Hamiltonians Eqs.~(\ref{H2a}) and (\ref{H2b}), the driving forces are well well-behaved for the parameter choices $\epsilon_{\rm GC}=50 \epsilon$ and $\sigma_{\rm GC}=0.735 \sigma$. Using the same MD parameters as those employed for the first thermodynamic path, the process linking the LJ system and the ideal gas through the GC model gives an excess Helmholtz free energy of
\begin{equation*}
\frac{\beta F^{\rm exc}_{\rm LJ}}{N}=(-0.10823 \pm 0.00005),
\end{equation*}
which, within the error bars, is identical to the result obtained using the sUF model as a reference system.

\subsection{Stillinger-Weber liquid Si}

As a second benchmark application we compute the excess Helmholtz free energy of a liquid-phase state of the Stillinger-Weber potential for Si,~\cite{Stillinger1985} which, in contrast to the LJ model, features strong directional bonding. The system was composed of 864 atoms in a periodic simulation cell at a number density of $5 \times 10^{-2}$~\AA$^{-3}$ and a temperature of 2600 $K$. The cutoff distance radius was $4.0 ~\AA$, the temperature is controlled using a Langevin thermostat and the equations of motion were integrated using the velocity Verlet algorithm with a timestep of $\Delta t = 0.5$~fs and a damping times scale of 50 fs. The equilibration and switching times used were $t_{eq} = 0.25$ns and $t_{sw} = 15$ ns, and we employ a linear form for the switching function $\lambda(t)$. In all cases, the average dynamical work estimators for both directions were determined from 10 independent realizations.

\begin{figure} [h!]
\includegraphics[width = 8.0 cm, angle=0 ]{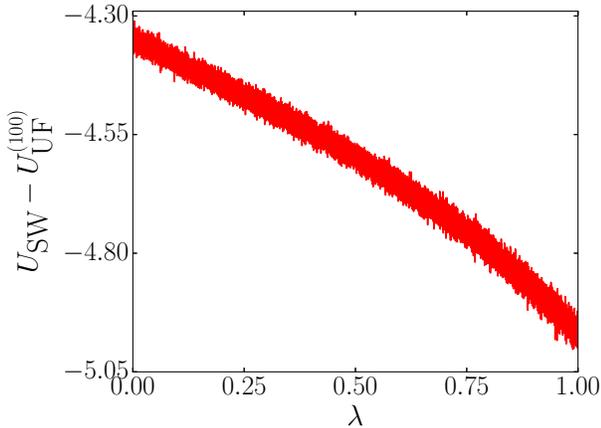}
\caption{Driving force $U_{\rm SW}-U^{(p)}_{\rm UF}$ as a function of $\lambda$ for $p=100$.}
\label{Fig9}
\end{figure}

We employ two thermodynamic paths analogous to those used for the LJ system to compute the excess Helmholtz free energy of the SW Si liquid state point. For the first path, analogous to that defined in Eq.~(\ref{H1}), we use the sUF model with $p=100$ and a length scale of $\sigma=1.3197602$~\AA, which corresponds to a value of $x=0.32$. A typical behavior of the driving force along the process is shown in Fig.~\ref{Fig9}. The average driving force is a slowly varying function and fluctuations are small and essentially independent of $\lambda$.   

Using the 10 forward and backward dynamical estimators to determine the free-energy difference between the SW and the UF models, and utilizing the cubic-spline-based expression for $F^{\rm exc (100)}_{\rm UF}(x)$, we obtain
\begin{equation*}
\frac{F^{\rm exc}_{\rm SW}}{N}=(-2.94909 \pm 0.00005)~\mbox{eV}.
\end{equation*}

Utilizing the analogous two-stage thermodynamic path defined by Eqs.~(\ref{H2a}) and (\ref{H2b}) for the LJ system, we determine a second, independent estimate for the SW excess Helmholtz free energy at the specified conditions. For the intermediate GC reference we employ the parameters $\varepsilon_{\rm GC}=50$~eV and $\sigma_{\rm GC}=1.08465$~\AA. Based on the 10 forward and backward realizations of both parts of the thermodynamic path we finally obtain
\begin{equation*}
\frac{F^{\rm exc}_{\rm SW}}{N}=(-2.94913 \pm 0.00006)~\mbox{eV}.
\end{equation*}
 
The two results are identical within the specified error bars, again demonstrating the usefulness of the sUF model as a reference system for fluid-phase free-energy calculations. 
 
\section{Summary}
\label{Sec5}
The UF potential was originally proposed as a model permitting the exact evaluation of its virial coefficients. In the present paper, in addition to computing three new virial coefficients, we assess its applicability as a reference system in fluid-phase free-energy calculations using molecular simulation. 

The UF model and its integer-scaled sUF variants possess a number of characteristics that render it an promising choice as a reference system in fluid-phase free-energy calculations. First, the particular temperature dependence of its potential-energy expression gives rise to thermodynamic properties in which the temperature plays the role of a mere scaling factor. In addition, their equations of state and excess Helmholtz free-energy expressions can be expressed in a scalable, adimensional form. In particular, both are fully characterized by a single function of one adimensional variable comprising the number density and interaction length scale. As a consequence, the models can be used as fluid-phase reference systems for any choice of temperature, number density and interaction length scale. 

In principle, these functions can be represented in terms of the virial coefficients that describe the equations of state. However, even though these coefficients can be determined exactly, in principle, their evaluation in practice is limited by the rapidly increasing amount of required graph-related computations. As a result, the number of known virial coefficients is limited and we find that truncated virial expansions are unable to provide a satisfactory precision for the description of the system's excess free-energies. To improve the numerical accuracy, we develop cubic-spline representations for the equations of state as well as the associated expressions for the excess Helmholtz free energies for 5 variants of the sUF model. They provide very accurate numerical representations of extensive MD data and all of them are available in the form of \verb python ~code in the Supplemental Material. 

Finally, we apply the UF and sUF models and the associated excess free-energy expressions as a reference system in the calculation of the excess Helmholtz free energy of a LJ fluid and a liquid state for the SW model for Si. Comparison of these results to those obtained from independent calculations demonstrates that, although the UF model is too soft, the sUF models offer a class of robust reference systems that permit highly accurate fluid-phase excess free-energy calculations without the need for an intermediate reference model. Indeed, in addition to the accuracy with which their free energies are known and the models' convenient scaling properties, for $p\lesssim 100$ the fluid is the only stable phase of the sUF model for any density and interaction length scale. This favorable set of properties may potentially put the sUF fluid-phase reference systems on par with the standard role that harmonic and Einstein solids play as references in solid-phase free-energy calculations. 

\section{Supplementary Material}

See supplementary material for \verb python ~script \verb ufGenerator.py ~that evaluates equation of state and excess Helmholtz free energy values for the UF and sUF models.  
\section*{Acknowledgments}

We gratefully acknowledge support from the Brazilian agencies CNPq, Fapesp, Capes and the Center for Computational Engineering and Sciences - Fapesp/Cepid no. 2013/08293-7. Part of the calculations were performed at CCJDR-IFGW-UNICAMP and CENAPAD-SP.

\bibliographystyle{apsrev4-1}
%\bibliography{/Users/dekoning/Documents/ReferenceDatabase/References} % Produces the bibliography via BibTeX.

\begin{thebibliography}{0}%
\makeatletter
\providecommand \@ifxundefined [1]{%
 \@ifx{#1\undefined}
}%
\providecommand \@ifnum [1]{%
 \ifnum #1\expandafter \@firstoftwo
 \else \expandafter \@secondoftwo
 \fi
}%
\providecommand \@ifx [1]{%
 \ifx #1\expandafter \@firstoftwo
 \else \expandafter \@secondoftwo
 \fi
}%
\providecommand \natexlab [1]{#1}%
\providecommand \enquote  [1]{``#1''}%
\providecommand \bibnamefont  [1]{#1}%
\providecommand \bibfnamefont [1]{#1}%
\providecommand \citenamefont [1]{#1}%
\providecommand \href@noop [0]{\@secondoftwo}%
\providecommand \href [0]{\begingroup \@sanitize@url \@href}%
\providecommand \@href[1]{\@@startlink{#1}\@@href}%
\providecommand \@@href[1]{\endgroup#1\@@endlink}%
\providecommand \@sanitize@url [0]{\catcode `\\12\catcode `\$12\catcode
  `\&12\catcode `\#12\catcode `\^12\catcode `\_12\catcode `\%12\relax}%
\providecommand \@@startlink[1]{}%
\providecommand \@@endlink[0]{}%
\providecommand \url  [0]{\begingroup\@sanitize@url \@url }%
\providecommand \@url [1]{\endgroup\@href {#1}{\urlprefix }}%
\providecommand \urlprefix  [0]{URL }%
\providecommand \Eprint [0]{\href }%
\providecommand \doibase [0]{http://dx.doi.org/}%
\providecommand \selectlanguage [0]{\@gobble}%
\providecommand \bibinfo  [0]{\@secondoftwo}%
\providecommand \bibfield  [0]{\@secondoftwo}%
\providecommand \translation [1]{[#1]}%
\providecommand \BibitemOpen [0]{}%
\providecommand \bibitemStop [0]{}%
\providecommand \bibitemNoStop [0]{.\EOS\space}%
\providecommand \EOS [0]{\spacefactor3000\relax}%
\providecommand \BibitemShut  [1]{\csname bibitem#1\endcsname}%
\let\auto@bib@innerbib\@empty
%</preamble>
\end{thebibliography}%


\begin{thebibliography}{52}%
\makeatletter
\providecommand \@ifxundefined [1]{%
 \@ifx{#1\undefined}
}%
\providecommand \@ifnum [1]{%
 \ifnum #1\expandafter \@firstoftwo
 \else \expandafter \@secondoftwo
 \fi
}%
\providecommand \@ifx [1]{%
 \ifx #1\expandafter \@firstoftwo
 \else \expandafter \@secondoftwo
 \fi
}%
\providecommand \natexlab [1]{#1}%
\providecommand \enquote  [1]{``#1''}%
\providecommand \bibnamefont  [1]{#1}%
\providecommand \bibfnamefont [1]{#1}%
\providecommand \citenamefont [1]{#1}%
\providecommand \href@noop [0]{\@secondoftwo}%
\providecommand \href [0]{\begingroup \@sanitize@url \@href}%
\providecommand \@href[1]{\@@startlink{#1}\@@href}%
\providecommand \@@href[1]{\endgroup#1\@@endlink}%
\providecommand \@sanitize@url [0]{\catcode `\\12\catcode `\$12\catcode
  `\&12\catcode `\#12\catcode `\^12\catcode `\_12\catcode `\%12\relax}%
\providecommand \@@startlink[1]{}%
\providecommand \@@endlink[0]{}%
\providecommand \url  [0]{\begingroup\@sanitize@url \@url }%
\providecommand \@url [1]{\endgroup\@href {#1}{\urlprefix }}%
\providecommand \urlprefix  [0]{URL }%
\providecommand \Eprint [0]{\href }%
\providecommand \doibase [0]{http://dx.doi.org/}%
\providecommand \selectlanguage [0]{\@gobble}%
\providecommand \bibinfo  [0]{\@secondoftwo}%
\providecommand \bibfield  [0]{\@secondoftwo}%
\providecommand \translation [1]{[#1]}%
\providecommand \BibitemOpen [0]{}%
\providecommand \bibitemStop [0]{}%
\providecommand \bibitemNoStop [0]{.\EOS\space}%
\providecommand \EOS [0]{\spacefactor3000\relax}%
\providecommand \BibitemShut  [1]{\csname bibitem#1\endcsname}%
\let\auto@bib@innerbib\@empty
%</preamble>
\bibitem [{\citenamefont {Mayer}\ and\ \citenamefont
  {Mayer}(1940)}]{Mayer1940}%
  \BibitemOpen
  \bibfield  {author} {\bibinfo {author} {\bibfnamefont {J.~E.}\ \bibnamefont
  {Mayer}}\ and\ \bibinfo {author} {\bibfnamefont {M.~G.}\ \bibnamefont
  {Mayer}},\ }\href {http://books.google.com.br/books?id=UsopZ_0iMvUC} {\emph
  {\bibinfo {title} {Statistical Mechanics}}}\ (\bibinfo  {publisher} {Wiley},\
  \bibinfo {year} {1940})\BibitemShut {NoStop}%
\bibitem [{\citenamefont {Mason}\ and\ \citenamefont
  {Spurling}(1969)}]{Mason1969}%
  \BibitemOpen
  \bibfield  {author} {\bibinfo {author} {\bibfnamefont {E.}~\bibnamefont
  {Mason}}\ and\ \bibinfo {author} {\bibfnamefont {T.}~\bibnamefont
  {Spurling}},\ }\href {https://books.google.com.br/books?id=c_INAQAAIAAJ}
  {\emph {\bibinfo {title} {The virial equation of state}}},\ \bibinfo {number}
  {v. 2}\ (\bibinfo  {publisher} {Pergamon Press},\ \bibinfo {year}
  {1969})\BibitemShut {NoStop}%
\bibitem [{\citenamefont {McQuarrie}(2000)}]{McQuarrie2000}%
  \BibitemOpen
  \bibfield  {author} {\bibinfo {author} {\bibfnamefont {D.}~\bibnamefont
  {McQuarrie}},\ }\href {http://books.google.com.br/books?id=itcpPnDnJM0C}
  {\emph {\bibinfo {title} {Statistical Mechanics}}}\ (\bibinfo  {publisher}
  {University Science Books},\ \bibinfo {year} {2000})\BibitemShut {NoStop}%
\bibitem [{\citenamefont {Hansen}\ and\ \citenamefont
  {McDonald}(2006)}]{Hansen2006}%
  \BibitemOpen
  \bibfield  {author} {\bibinfo {author} {\bibfnamefont {J.}~\bibnamefont
  {Hansen}}\ and\ \bibinfo {author} {\bibfnamefont {I.}~\bibnamefont
  {McDonald}},\ }\href {http://books.google.com/books?id=Uhm87WZBnxEC} {\emph
  {\bibinfo {title} {Theory of simple liquids}}},\ \bibinfo {edition} {3rd}\
  ed.\ (\bibinfo  {publisher} {Elsevier Academic Press},\ \bibinfo {year}
  {2006})\BibitemShut {NoStop}%
\bibitem [{\citenamefont {Groeneveld}(1962)}]{Groeneveld1962}%
  \BibitemOpen
  \bibfield  {author} {\bibinfo {author} {\bibfnamefont {J.}~\bibnamefont
  {Groeneveld}},\ }\href
  {http://www.sciencedirect.com/science/article/pii/0031916362901981}
  {\bibfield  {journal} {\bibinfo  {journal} {Phys. Lett.}\ }\textbf {\bibinfo
  {volume} {3}},\ \bibinfo {pages} {50} (\bibinfo {year} {1962})}\BibitemShut
  {NoStop}%
\bibitem [{\citenamefont {Baram}\ and\ \citenamefont
  {Rowlinson}(1990)}]{Baram1990}%
  \BibitemOpen
  \bibfield  {author} {\bibinfo {author} {\bibfnamefont {A.}~\bibnamefont
  {Baram}}\ and\ \bibinfo {author} {\bibfnamefont {J.~S.}\ \bibnamefont
  {Rowlinson}},\ }\href {http://stacks.iop.org/0305-4470/23/i=8/a=009}
  {\bibfield  {journal} {\bibinfo  {journal} {J. Phys. A}\ }\textbf {\bibinfo
  {volume} {23}},\ \bibinfo {pages} {L399} (\bibinfo {year}
  {1990})}\BibitemShut {NoStop}%
\bibitem [{\citenamefont {Baram}\ and\ \citenamefont
  {Rowlinson}(1991)}]{Baram1991}%
  \BibitemOpen
  \bibfield  {author} {\bibinfo {author} {\bibfnamefont {A.}~\bibnamefont
  {Baram}}\ and\ \bibinfo {author} {\bibfnamefont {J.~S.}\ \bibnamefont
  {Rowlinson}},\ }\href {\doibase 10.1080/00268979100102521} {\bibfield
  {journal} {\bibinfo  {journal} {Mol. Phys.}\ }\textbf {\bibinfo {volume}
  {74}},\ \bibinfo {pages} {707} (\bibinfo {year} {1991})}\BibitemShut
  {NoStop}%
\bibitem [{\citenamefont {Clisby}\ and\ \citenamefont
  {McCoy}(2006)}]{Clisby2006}%
  \BibitemOpen
  \bibfield  {author} {\bibinfo {author} {\bibfnamefont {N.}~\bibnamefont
  {Clisby}}\ and\ \bibinfo {author} {\bibfnamefont {B.~M.}\ \bibnamefont
  {McCoy}},\ }\href {http://dx.doi.org/10.1007/s10955-005-8080-0} {\bibfield
  {journal} {\bibinfo  {journal} {J. Stat. Phys.}\ }\textbf {\bibinfo {volume}
  {122}},\ \bibinfo {pages} {15} (\bibinfo {year} {2006})}\BibitemShut
  {NoStop}%
\bibitem [{\citenamefont {Eisenberg}\ and\ \citenamefont
  {Baram}(2007)}]{Eisenberg2007}%
  \BibitemOpen
  \bibfield  {author} {\bibinfo {author} {\bibfnamefont {E.}~\bibnamefont
  {Eisenberg}}\ and\ \bibinfo {author} {\bibfnamefont {A.}~\bibnamefont
  {Baram}},\ }\href {\doibase 10.1073/pnas.0700778104} {\bibfield  {journal}
  {\bibinfo  {journal} {Proc. Natl. Acad. Sci. U.S.A.}\ }\textbf {\bibinfo
  {volume} {104}},\ \bibinfo {pages} {5755} (\bibinfo {year}
  {2007})}\BibitemShut {NoStop}%
\bibitem [{\citenamefont {Ushcats}(2012)}]{Ushcats2012}%
  \BibitemOpen
  \bibfield  {author} {\bibinfo {author} {\bibfnamefont {M.~V.}\ \bibnamefont
  {Ushcats}},\ }\href {http://link.aps.org/doi/10.1103/PhysRevLett.109.040601}
  {\bibfield  {journal} {\bibinfo  {journal} {Phys. Rev. Lett.}\ }\textbf
  {\bibinfo {volume} {109}},\ \bibinfo {pages} {040601} (\bibinfo {year}
  {2012})}\BibitemShut {NoStop}%
\bibitem [{\citenamefont {Bannur}(2015)}]{Bannur2015}%
  \BibitemOpen
  \bibfield  {author} {\bibinfo {author} {\bibfnamefont {V.~M.}\ \bibnamefont
  {Bannur}},\ }\href
  {http://www.sciencedirect.com/science/article/pii/S0378437114008917}
  {\bibfield  {journal} {\bibinfo  {journal} {Physica A: Statistical Mechanics
  and its Applications}\ }\textbf {\bibinfo {volume} {419}},\ \bibinfo {pages}
  {675} (\bibinfo {year} {2015})}\BibitemShut {NoStop}%
\bibitem [{\citenamefont {Lab{\'i}k}\ \emph {et~al.}(2005)\citenamefont
  {Lab{\'i}k}, \citenamefont {Kolafa},\ and\ \citenamefont
  {Malijevsk{\'y}}}]{Labik2005}%
  \BibitemOpen
  \bibfield  {author} {\bibinfo {author} {\bibfnamefont {S.}~\bibnamefont
  {Lab{\'i}k}}, \bibinfo {author} {\bibfnamefont {J.}~\bibnamefont {Kolafa}}, \
  and\ \bibinfo {author} {\bibfnamefont {A.}~\bibnamefont {Malijevsk{\'y}}},\
  }\href {http://link.aps.org/doi/10.1103/PhysRevE.71.021105} {\bibfield
  {journal} {\bibinfo  {journal} {Phys. Rev. E}\ }\textbf {\bibinfo {volume}
  {71}},\ \bibinfo {pages} {021105} (\bibinfo {year} {2005})}\BibitemShut
  {NoStop}%
\bibitem [{\citenamefont {Masters}(2008)}]{Masters2008}%
  \BibitemOpen
  \bibfield  {author} {\bibinfo {author} {\bibfnamefont {A.~J.}\ \bibnamefont
  {Masters}},\ }\href {http://stacks.iop.org/0953-8984/20/i=28/a=283102}
  {\bibfield  {journal} {\bibinfo  {journal} {J. Phys.: Condens. Matter}\
  }\textbf {\bibinfo {volume} {20}},\ \bibinfo {pages} {283102} (\bibinfo
  {year} {2008})}\BibitemShut {NoStop}%
\bibitem [{\citenamefont {Wheatley}(2013)}]{Wheatley2013}%
  \BibitemOpen
  \bibfield  {author} {\bibinfo {author} {\bibfnamefont {R.~J.}\ \bibnamefont
  {Wheatley}},\ }\href {http://link.aps.org/doi/10.1103/PhysRevLett.110.200601}
  {\bibfield  {journal} {\bibinfo  {journal} {Phys. Rev. Lett.}\ }\textbf
  {\bibinfo {volume} {110}},\ \bibinfo {pages} {200601} (\bibinfo {year}
  {2013})}\BibitemShut {NoStop}%
\bibitem [{\citenamefont {Schultz}\ and\ \citenamefont
  {Kofke}(2014)}]{Schultz2014}%
  \BibitemOpen
  \bibfield  {author} {\bibinfo {author} {\bibfnamefont {A.~J.}\ \bibnamefont
  {Schultz}}\ and\ \bibinfo {author} {\bibfnamefont {D.~A.}\ \bibnamefont
  {Kofke}},\ }\href {http://link.aps.org/doi/10.1103/PhysRevE.90.023301}
  {\bibfield  {journal} {\bibinfo  {journal} {Phys. Rev. E}\ }\textbf {\bibinfo
  {volume} {90}},\ \bibinfo {pages} {023301} (\bibinfo {year}
  {2014})}\BibitemShut {NoStop}%
\bibitem [{\citenamefont {Wheatley}(2005)}]{Wheatley2005}%
  \BibitemOpen
  \bibfield  {author} {\bibinfo {author} {\bibfnamefont {R.~J.}\ \bibnamefont
  {Wheatley}},\ }\href {\doibase 10.1021/jp040709i} {\bibfield  {journal}
  {\bibinfo  {journal} {J. Phys. Chem. B}\ }\textbf {\bibinfo {volume} {109}},\
  \bibinfo {pages} {7463} (\bibinfo {year} {2005})}\BibitemShut {NoStop}%
\bibitem [{\citenamefont {Tan}\ \emph {et~al.}(2010)\citenamefont {Tan},
  \citenamefont {Schultz},\ and\ \citenamefont {Kofke}}]{Tan2010}%
  \BibitemOpen
  \bibfield  {author} {\bibinfo {author} {\bibfnamefont {T.~B.}\ \bibnamefont
  {Tan}}, \bibinfo {author} {\bibfnamefont {A.~J.}\ \bibnamefont {Schultz}}, \
  and\ \bibinfo {author} {\bibfnamefont {D.~A.}\ \bibnamefont {Kofke}},\ }\href
  {\doibase 10.1080/00268976.2010.520041} {\bibfield  {journal} {\bibinfo
  {journal} {Mol. Phys.}\ }\textbf {\bibinfo {volume} {109}},\ \bibinfo {pages}
  {123} (\bibinfo {year} {2010})}\BibitemShut {NoStop}%
\bibitem [{\citenamefont {Singh}\ and\ \citenamefont
  {Kofke}(2004)}]{Singh2004}%
  \BibitemOpen
  \bibfield  {author} {\bibinfo {author} {\bibfnamefont {J.~K.}\ \bibnamefont
  {Singh}}\ and\ \bibinfo {author} {\bibfnamefont {D.~A.}\ \bibnamefont
  {Kofke}},\ }\href {\doibase 10.1103/PhysRevLett.92.220601} {\bibfield
  {journal} {\bibinfo  {journal} {Phys. Rev. Lett.}\ }\textbf {\bibinfo
  {volume} {92}},\ \bibinfo {pages} {220601} (\bibinfo {year}
  {2004})}\BibitemShut {NoStop}%
\bibitem [{\citenamefont {Schultz}\ and\ \citenamefont
  {Kofke}(2009)}]{Schultz2009}%
  \BibitemOpen
  \bibfield  {author} {\bibinfo {author} {\bibfnamefont {A.~J.}\ \bibnamefont
  {Schultz}}\ and\ \bibinfo {author} {\bibfnamefont {D.~A.}\ \bibnamefont
  {Kofke}},\ }\href {\doibase 10.1080/00268970903267053} {\bibfield  {journal}
  {\bibinfo  {journal} {Mol. Phys.}\ }\textbf {\bibinfo {volume} {107}},\
  \bibinfo {pages} {2309} (\bibinfo {year} {2009})}\BibitemShut {NoStop}%
\bibitem [{\citenamefont {Schultz}\ \emph {et~al.}(2012)\citenamefont
  {Schultz}, \citenamefont {Barlow}, \citenamefont {Chaudhary},\ and\
  \citenamefont {Kofke}}]{Schultz2012}%
  \BibitemOpen
  \bibfield  {author} {\bibinfo {author} {\bibfnamefont {A.~J.}\ \bibnamefont
  {Schultz}}, \bibinfo {author} {\bibfnamefont {N.~S.}\ \bibnamefont {Barlow}},
  \bibinfo {author} {\bibfnamefont {V.}~\bibnamefont {Chaudhary}}, \ and\
  \bibinfo {author} {\bibfnamefont {D.~A.}\ \bibnamefont {Kofke}},\ }\href
  {\doibase 10.1080/00268976.2012.730642} {\bibfield  {journal} {\bibinfo
  {journal} {Mol. Phys.}\ }\textbf {\bibinfo {volume} {111}},\ \bibinfo {pages}
  {535} (\bibinfo {year} {2012})}\BibitemShut {NoStop}%
\bibitem [{\citenamefont {Ushcats}(2014)}]{Ushcats2014}%
  \BibitemOpen
  \bibfield  {author} {\bibinfo {author} {\bibfnamefont {M.~V.}\ \bibnamefont
  {Ushcats}},\ }\href {\doibase http://dx.doi.org/10.1063/1.4895126} {\bibfield
   {journal} {\bibinfo  {journal} {J. Chem. Phys.}\ }\textbf {\bibinfo {volume}
  {141}},\ \bibinfo {pages} {101103} (\bibinfo {year} {2014})}\BibitemShut
  {NoStop}%
\bibitem [{\citenamefont {Feng}\ \emph {et~al.}(2015)\citenamefont {Feng},
  \citenamefont {Schultz}, \citenamefont {Chaudhary},\ and\ \citenamefont
  {Kofke}}]{Feng2015}%
  \BibitemOpen
  \bibfield  {author} {\bibinfo {author} {\bibfnamefont {C.}~\bibnamefont
  {Feng}}, \bibinfo {author} {\bibfnamefont {A.~J.}\ \bibnamefont {Schultz}},
  \bibinfo {author} {\bibfnamefont {V.}~\bibnamefont {Chaudhary}}, \ and\
  \bibinfo {author} {\bibfnamefont {D.~A.}\ \bibnamefont {Kofke}},\ }\href
  {\doibase http://dx.doi.org/10.1063/1.4927339} {\bibfield  {journal}
  {\bibinfo  {journal} {J. Chem. Phys.}\ }\textbf {\bibinfo {volume} {143}},\
  \bibinfo {eid} {044504} (\bibinfo {year} {2015})}\BibitemShut {NoStop}%
\bibitem [{\citenamefont {Allen}\ and\ \citenamefont
  {Tildesley}(1987)}]{AllenTildesley}%
  \BibitemOpen
  \bibfield  {author} {\bibinfo {author} {\bibfnamefont {M.~P.}\ \bibnamefont
  {Allen}}\ and\ \bibinfo {author} {\bibfnamefont {D.~J.}\ \bibnamefont
  {Tildesley}},\ }\href@noop {} {\emph {\bibinfo {title} {Computer Simulation
  of Liquids}}}\ (\bibinfo  {publisher} {Oxford University Press},\ \bibinfo
  {year} {1987})\BibitemShut {NoStop}%
\bibitem [{\citenamefont {Frenkel}\ and\ \citenamefont
  {Smit}(2002)}]{Frenkel2002}%
  \BibitemOpen
  \bibfield  {author} {\bibinfo {author} {\bibfnamefont {D.}~\bibnamefont
  {Frenkel}}\ and\ \bibinfo {author} {\bibfnamefont {B.}~\bibnamefont {Smit}},\
  }\href@noop {} {\emph {\bibinfo {title} {Understanding Molecular
  Simulation}}}\ (\bibinfo  {publisher} {Academic Press, San Diego},\ \bibinfo
  {address} {San Diego},\ \bibinfo {year} {2002})\BibitemShut {NoStop}%
\bibitem [{\citenamefont {Chipot}\ and\ \citenamefont
  {Pohorille}(2007)}]{Chipot2007}%
  \BibitemOpen
  \bibfield  {author} {\bibinfo {author} {\bibfnamefont {C.}~\bibnamefont
  {Chipot}}\ and\ \bibinfo {author} {\bibfnamefont {A.}~\bibnamefont
  {Pohorille}},\ }\href {https://books.google.com.br/books?id=XTE5eejRTC0C}
  {\emph {\bibinfo {title} {Free Energy Calculations: Theory and Applications
  in Chemistry and Biology}}}\ (\bibinfo  {publisher} {Springer Berlin
  Heidelberg},\ \bibinfo {year} {2007})\BibitemShut {NoStop}%
\bibitem [{\citenamefont {Leli{\`e}vre}\ \emph {et~al.}(2010)\citenamefont
  {Leli{\`e}vre}, \citenamefont {Stoltz},\ and\ \citenamefont
  {Rousset}}]{Lelievre2010}%
  \BibitemOpen
  \bibfield  {author} {\bibinfo {author} {\bibfnamefont {T.}~\bibnamefont
  {Leli{\`e}vre}}, \bibinfo {author} {\bibfnamefont {G.}~\bibnamefont
  {Stoltz}}, \ and\ \bibinfo {author} {\bibfnamefont {M.}~\bibnamefont
  {Rousset}},\ }\href {https://books.google.com.br/books?id=SqJGgfPq_ZUC}
  {\emph {\bibinfo {title} {Free Energy Computations: A Mathematical
  Perspective}}}\ (\bibinfo  {publisher} {Imperial College Press},\ \bibinfo
  {year} {2010})\BibitemShut {NoStop}%
\bibitem [{\citenamefont {Abramo}\ \emph {et~al.}(2015)\citenamefont {Abramo},
  \citenamefont {Caccamo}, \citenamefont {Costa}, \citenamefont {Giaquinta},
  \citenamefont {Malescio}, \citenamefont {Muna{\`o}},\ and\ \citenamefont
  {Prestipino}}]{Abramo2015}%
  \BibitemOpen
  \bibfield  {author} {\bibinfo {author} {\bibfnamefont {M.~C.}\ \bibnamefont
  {Abramo}}, \bibinfo {author} {\bibfnamefont {C.}~\bibnamefont {Caccamo}},
  \bibinfo {author} {\bibfnamefont {D.}~\bibnamefont {Costa}}, \bibinfo
  {author} {\bibfnamefont {P.~V.}\ \bibnamefont {Giaquinta}}, \bibinfo {author}
  {\bibfnamefont {G.}~\bibnamefont {Malescio}}, \bibinfo {author}
  {\bibfnamefont {G.}~\bibnamefont {Muna{\`o}}}, \ and\ \bibinfo {author}
  {\bibfnamefont {S.}~\bibnamefont {Prestipino}},\ }\href {\doibase
  http://dx.doi.org/10.1063/1.4921884} {\bibfield  {journal} {\bibinfo
  {journal} {J. Chem. Phys.}\ }\textbf {\bibinfo {volume}
  {142}},\ \bibinfo {pages} {214502} (\bibinfo {year} {2015})}\BibitemShut
  {NoStop}%
\bibitem [{\citenamefont {Broughton}\ and\ \citenamefont
  {Gilmer}(1983)}]{Broughton1983I}%
  \BibitemOpen
  \bibfield  {author} {\bibinfo {author} {\bibfnamefont {J.~Q.}\ \bibnamefont
  {Broughton}}\ and\ \bibinfo {author} {\bibfnamefont {G.~H.}\ \bibnamefont
  {Gilmer}},\ }\href {\doibase http://dx.doi.org/10.1063/1.445633} {\bibfield
  {journal} {\bibinfo  {journal} {J. Chem. Phys.}\ }\textbf {\bibinfo {volume}
  {79}},\ \bibinfo {pages} {5095} (\bibinfo {year} {1983})}\BibitemShut
  {NoStop}%
\bibitem [{\citenamefont {Broughton}\ and\ \citenamefont
  {Li}(1987)}]{Broughton1987}%
  \BibitemOpen
  \bibfield  {author} {\bibinfo {author} {\bibfnamefont {J.~Q.}\ \bibnamefont
  {Broughton}}\ and\ \bibinfo {author} {\bibfnamefont {X.~P.}\ \bibnamefont
  {Li}},\ }\href {http://link.aps.org/doi/10.1103/PhysRevB.35.9120} {\bibfield
  {journal} {\bibinfo  {journal} {Phys. Rev. B}\ }\textbf {\bibinfo {volume}
  {35}},\ \bibinfo {pages} {9120} (\bibinfo {year} {1987})}\BibitemShut
  {NoStop}%
\bibitem [{\citenamefont {van~der Waals}(1873)}]{vanderWaals1873}%
  \BibitemOpen
  \bibfield  {author} {\bibinfo {author} {\bibfnamefont {J.}~\bibnamefont
  {van~der Waals}},\ }\emph {\bibinfo {title} {Studies in statistical
  mechanics}},\ \href {https://books.google.com.br/books?id=FC5EAQAAIAAJ}
  {Ph.D. thesis},\ \bibinfo  {school} {Hoogeschool te Leiden} (\bibinfo {year}
  {1873}),\ \bibinfo {note} {engl. Transl. J. S. Rowlinson, Studies in
  Statistical Mechanics vol 14 (North-Holland, Amsterdam, 1988)}\BibitemShut
  {NoStop}%
\bibitem [{\citenamefont {Stillinger}(1976)}]{Stillinger1976}%
  \BibitemOpen
  \bibfield  {author} {\bibinfo {author} {\bibfnamefont {F.~H.}\ \bibnamefont
  {Stillinger}},\ }\href {\doibase http://dx.doi.org/10.1063/1.432891}
  {\bibfield  {journal} {\bibinfo  {journal} {J. Chem. Phys.}\ }\textbf
  {\bibinfo {volume} {65}},\ \bibinfo {pages} {3968} (\bibinfo {year}
  {1976})}\BibitemShut {NoStop}%
\bibitem [{\citenamefont {Prestipino}\ \emph
  {et~al.}(2005{\natexlab{a}})\citenamefont {Prestipino}, \citenamefont
  {Saija},\ and\ \citenamefont {Giaquinta}}]{Prestipino2005}%
  \BibitemOpen
  \bibfield  {author} {\bibinfo {author} {\bibfnamefont {S.}~\bibnamefont
  {Prestipino}}, \bibinfo {author} {\bibfnamefont {F.}~\bibnamefont {Saija}}, \
  and\ \bibinfo {author} {\bibfnamefont {P.~V.}\ \bibnamefont {Giaquinta}},\
  }\href {\doibase http://dx.doi.org/10.1063/1.2064639} {\bibfield  {journal}
  {\bibinfo  {journal} {J. Chem. Phys.}\ }\textbf {\bibinfo {volume} {123}},\
  \bibinfo {eid} {144110} (\bibinfo {year} {2005}{\natexlab{a}})}\BibitemShut
  {NoStop}%
\bibitem [{\citenamefont {Prestipino}\ \emph
  {et~al.}(2005{\natexlab{b}})\citenamefont {Prestipino}, \citenamefont
  {Saija},\ and\ \citenamefont {Giaquinta}}]{Prestipino2005a}%
  \BibitemOpen
  \bibfield  {author} {\bibinfo {author} {\bibfnamefont {S.}~\bibnamefont
  {Prestipino}}, \bibinfo {author} {\bibfnamefont {F.}~\bibnamefont {Saija}}, \
  and\ \bibinfo {author} {\bibfnamefont {P.~V.}\ \bibnamefont {Giaquinta}},\
  }\href {http://link.aps.org/doi/10.1103/PhysRevE.71.050102} {\bibfield
  {journal} {\bibinfo  {journal} {Phys. Rev. E}\ }\textbf {\bibinfo {volume}
  {71}},\ \bibinfo {pages} {050102} (\bibinfo {year}
  {2005}{\natexlab{b}})}\BibitemShut {NoStop}%
\bibitem [{\citenamefont {Ryu}\ and\ \citenamefont {Cai}(2008)}]{Ryu2008a}%
  \BibitemOpen
  \bibfield  {author} {\bibinfo {author} {\bibfnamefont {S.}~\bibnamefont
  {Ryu}}\ and\ \bibinfo {author} {\bibfnamefont {W.}~\bibnamefont {Cai}},\
  }\href {http://stacks.iop.org/0965-0393/16/i=8/a=085005} {\bibfield
  {journal} {\bibinfo  {journal} {Model. Simul. Mater. Sci. Eng.}\ }\textbf
  {\bibinfo {volume} {16}},\ \bibinfo {pages} {085005} (\bibinfo {year}
  {2008})}\BibitemShut {NoStop}%
\bibitem [{\citenamefont {Hoover}\ \emph {et~al.}(1970)\citenamefont {Hoover},
  \citenamefont {Ross}, \citenamefont {Johnson}, \citenamefont {Henderson},
  \citenamefont {Barker},\ and\ \citenamefont {Brown}}]{Hoover1970}%
  \BibitemOpen
  \bibfield  {author} {\bibinfo {author} {\bibfnamefont {W.~G.}\ \bibnamefont
  {Hoover}}, \bibinfo {author} {\bibfnamefont {M.}~\bibnamefont {Ross}},
  \bibinfo {author} {\bibfnamefont {K.~W.}\ \bibnamefont {Johnson}}, \bibinfo
  {author} {\bibfnamefont {D.}~\bibnamefont {Henderson}}, \bibinfo {author}
  {\bibfnamefont {J.~A.}\ \bibnamefont {Barker}}, \ and\ \bibinfo {author}
  {\bibfnamefont {B.~C.}\ \bibnamefont {Brown}},\ }\href {\doibase
  http://dx.doi.org/10.1063/1.1672728} {\bibfield  {journal} {\bibinfo
  {journal} {J. Chem. Phys.}\ }\textbf {\bibinfo {volume} {52}},\ \bibinfo
  {pages} {4931} (\bibinfo {year} {1970})}\BibitemShut {NoStop}%
\bibitem [{\citenamefont {Young}\ and\ \citenamefont
  {Rogers}(1984)}]{Young1984}%
  \BibitemOpen
  \bibfield  {author} {\bibinfo {author} {\bibfnamefont {D.~A.}\ \bibnamefont
  {Young}}\ and\ \bibinfo {author} {\bibfnamefont {F.~J.}\ \bibnamefont
  {Rogers}},\ }\href {\doibase http://dx.doi.org/10.1063/1.447951} {\bibfield
  {journal} {\bibinfo  {journal} {J. Chem. Phys.}\ }\textbf
  {\bibinfo {volume} {81}},\ \bibinfo {pages} {2789} (\bibinfo {year}
  {1984})}\BibitemShut {NoStop}%
\bibitem [{\citenamefont {de~Koning}\ \emph {et~al.}(2001)\citenamefont
  {de~Koning}, \citenamefont {Antonelli},\ and\ \citenamefont
  {Yip}}]{deKoning2001}%
  \BibitemOpen
  \bibfield  {author} {\bibinfo {author} {\bibfnamefont {M.}~\bibnamefont
  {de~Koning}}, \bibinfo {author} {\bibfnamefont {A.}~\bibnamefont
  {Antonelli}}, \ and\ \bibinfo {author} {\bibfnamefont {S.}~\bibnamefont
  {Yip}},\ }\href {http://dx.doi.org/10.1063/1.1420486} {\bibfield  {journal}
  {\bibinfo  {journal} {J. Chem. Phys.}\ }\textbf {\bibinfo {volume} {115}},\
  \bibinfo {pages} {11025} (\bibinfo {year} {2001})}\BibitemShut {NoStop}%
\bibitem [{\citenamefont {Michelon}\ and\ \citenamefont
  {Antonelli}(2010)}]{Michelon2010}%
  \BibitemOpen
  \bibfield  {author} {\bibinfo {author} {\bibfnamefont {M.~F.}\ \bibnamefont
  {Michelon}}\ and\ \bibinfo {author} {\bibfnamefont {A.}~\bibnamefont
  {Antonelli}},\ }\href {http://link.aps.org/doi/10.1103/PhysRevB.81.094204}
  {\bibfield  {journal} {\bibinfo  {journal} {Phys. Rev. B}\ }\textbf {\bibinfo
  {volume} {81}},\ \bibinfo {pages} {094204} (\bibinfo {year}
  {2010})}\BibitemShut {NoStop}%
\bibitem [{\citenamefont {Greeff}(2008)}]{Greeff2008}%
  \BibitemOpen
  \bibfield  {author} {\bibinfo {author} {\bibfnamefont {C.~W.}\ \bibnamefont
  {Greeff}},\ }\href {\doibase http://dx.doi.org/10.1063/1.2917355} {\bibfield
  {journal} {\bibinfo  {journal} {J. Chem. Phys.}\ }\textbf {\bibinfo {volume}
  {128}},\ \bibinfo {pages} {184104} (\bibinfo {year} {2008})}\BibitemShut
  {NoStop}%
\bibitem [{\citenamefont {Moriarty}\ and\ \citenamefont
  {Haskins}(2014)}]{Moriarty2014}%
  \BibitemOpen
  \bibfield  {author} {\bibinfo {author} {\bibfnamefont {J.~A.}\ \bibnamefont
  {Moriarty}}\ and\ \bibinfo {author} {\bibfnamefont {J.~B.}\ \bibnamefont
  {Haskins}},\ }\href {http://link.aps.org/doi/10.1103/PhysRevB.90.054113}
  {\bibfield  {journal} {\bibinfo  {journal} {Phys. Rev. B}\ }\textbf {\bibinfo
  {volume} {90}},\ \bibinfo {pages} {054113} (\bibinfo {year}
  {2014})}\BibitemShut {NoStop}%
\bibitem [{\citenamefont {Uhlenbeck}\ and\ \citenamefont
  {Ford}(1962)}]{deBoer1962}%
  \BibitemOpen
  \bibfield  {author} {\bibinfo {author} {\bibfnamefont {G.}~\bibnamefont
  {Uhlenbeck}}\ and\ \bibinfo {author} {\bibfnamefont {G.}~\bibnamefont
  {Ford}},\ }\enquote {\bibinfo {title} {Studies in statistical mechanics},}\
  in\ \href {http://books.google.com.br/books?id=ZDEPAAAAIAAJ} {\emph {\bibinfo
  {booktitle} {Series in physics}}},\ \bibinfo {series and number} {\bibinfo
  {number} {v. 1}},\ \bibinfo {editor} {edited by\ \bibinfo {editor}
  {\bibfnamefont {J.}~\bibnamefont {de~Boer}}\ and\ \bibinfo {editor}
  {\bibfnamefont {G.}~\bibnamefont {Uhlenbeck}}}\ (\bibinfo  {publisher}
  {North-Holland Publishing Company},\ \bibinfo {year} {1962})\ Chap.\ \bibinfo
  {chapter} {The development of linear graphs with applications to the theory
  of the virial development of the properties of gases}, p.\ \bibinfo {pages}
  {182}\BibitemShut {NoStop}%
\bibitem [{\citenamefont {Stillinger}\ and\ \citenamefont
  {Weber}(1985)}]{Stillinger1985}%
  \BibitemOpen
  \bibfield  {author} {\bibinfo {author} {\bibfnamefont {F.~H.}\ \bibnamefont
  {Stillinger}}\ and\ \bibinfo {author} {\bibfnamefont {T.~A.}\ \bibnamefont
  {Weber}},\ }\href {http://dx.doi.org/10.1103/PhysRevB.31.5262} {\bibfield
  {journal} {\bibinfo  {journal} {Phys. Rev. B}\ }\textbf {\bibinfo {volume}
  {31}},\ \bibinfo {pages} {5262} (\bibinfo {year} {1985})}\BibitemShut
  {NoStop}%
\bibitem [{\citenamefont {Feynman}(1998)}]{Feynman1998}%
  \BibitemOpen
  \bibfield  {author} {\bibinfo {author} {\bibfnamefont {R.}~\bibnamefont
  {Feynman}},\ }\href {http://books.google.com.br/books?id=Ou4ltPYiXPgC} {\emph
  {\bibinfo {title} {Statistical Mechanics: A Set Of Lectures}}}\ (\bibinfo
  {publisher} {Westview Press},\ \bibinfo {year} {1998})\BibitemShut {NoStop}%
\bibitem [{\citenamefont {Newman}(2010)}]{Newman2010}%
  \BibitemOpen
  \bibfield  {author} {\bibinfo {author} {\bibfnamefont {M.~E.~J.}\
  \bibnamefont {Newman}},\ }\href
  {https://books.google.com.br/books?id=q7HVtpYVfC0C} {\emph {\bibinfo {title}
  {Networks: An Introduction}}}\ (\bibinfo  {publisher} {Oxford University
  Press, Oxford},\ \bibinfo {year} {2010})\BibitemShut {NoStop}%
\bibitem [{\citenamefont {McKay}\ and\ \citenamefont
  {Piperno}(2014)}]{McKay2014}%
  \BibitemOpen
  \bibfield  {author} {\bibinfo {author} {\bibfnamefont {B.~D.}\ \bibnamefont
  {McKay}}\ and\ \bibinfo {author} {\bibfnamefont {A.}~\bibnamefont
  {Piperno}},\ }\href
  {http://www.sciencedirect.com/science/article/pii/S0747717113001193}
  {\bibfield  {journal} {\bibinfo  {journal} {J. Symb. Comput.}\
  }\textbf {\bibinfo {volume} {60}},\ \bibinfo {pages} {94} (\bibinfo {year}
  {2014})}\BibitemShut {NoStop}%
\bibitem [{\citenamefont {Bareiss}(1968)}]{Bareiss1968}%
  \BibitemOpen
  \bibfield  {author} {\bibinfo {author} {\bibfnamefont {E.~H.}\ \bibnamefont
  {Bareiss}},\ }\href {http://www.jstor.org/stable/2004533} {\bibfield
  {journal} {\bibinfo  {journal} {Math. Comp.}\ }\textbf {\bibinfo {volume}
  {22}},\ \bibinfo {pages} {565} (\bibinfo {year} {1968})}\BibitemShut
  {NoStop}%
\bibitem [{\citenamefont {Plimpton}(1995)}]{Plimpton1995}%
  \BibitemOpen
  \bibfield  {author} {\bibinfo {author} {\bibfnamefont {S.}~\bibnamefont
  {Plimpton}},\ }\href
  {http://www.sciencedirect.com/science/article/B6WHY-45NJN1B-3N/2/58aa2a309d2ebbbe60e0f417d398b0ef}
  {\bibfield  {journal} {\bibinfo  {journal} {J. Comput. Phys.}\ }\textbf
  {\bibinfo {volume} {117}},\ \bibinfo {pages} {1} (\bibinfo {year}
  {1995})}\BibitemShut {NoStop}%
\bibitem [{\citenamefont {Press}\ \emph {et~al.}(2007)\citenamefont {Press},
  \citenamefont {Teukolsky}, \citenamefont {Vetterling},\ and\ \citenamefont
  {Flannery}}]{Press2007}%
  \BibitemOpen
  \bibfield  {author} {\bibinfo {author} {\bibfnamefont {W.}~\bibnamefont
  {Press}}, \bibinfo {author} {\bibfnamefont {S.}~\bibnamefont {Teukolsky}},
  \bibinfo {author} {\bibfnamefont {W.}~\bibnamefont {Vetterling}}, \ and\
  \bibinfo {author} {\bibfnamefont {B.}~\bibnamefont {Flannery}},\ }\href
  {https://books.google.com.br/books?id=1aAOdzK3FegC} {\emph {\bibinfo {title}
  {Numerical Recipes 3rd Edition: The Art of Scientific Computing}}}\ (\bibinfo
   {publisher} {Cambridge University Press},\ \bibinfo {year}
  {2007})\BibitemShut {NoStop}%
\bibitem [{\citenamefont {Watanabe}\ and\ \citenamefont
  {Reinhardt}(1990)}]{Watanabe1990}%
  \BibitemOpen
  \bibfield  {author} {\bibinfo {author} {\bibfnamefont {M.}~\bibnamefont
  {Watanabe}}\ and\ \bibinfo {author} {\bibfnamefont {W.~P.}\ \bibnamefont
  {Reinhardt}},\ }\href {http://link.aps.org/doi/10.1103/PhysRevLett.65.3301}
  {\bibfield  {journal} {\bibinfo  {journal} {Phys. Rev. Lett.}\ }\textbf
  {\bibinfo {volume} {65}},\ \bibinfo {pages} {3301} (\bibinfo {year}
  {1990})}\BibitemShut {NoStop}%
\bibitem [{\citenamefont {de~Koning}\ and\ \citenamefont
  {Antonelli}(1997)}]{deKoning1997}%
  \BibitemOpen
  \bibfield  {author} {\bibinfo {author} {\bibfnamefont {M.}~\bibnamefont
  {de~Koning}}\ and\ \bibinfo {author} {\bibfnamefont {A.}~\bibnamefont
  {Antonelli}},\ }\href {http://link.aps.org/doi/10.1103/PhysRevB.55.735}
  {\bibfield  {journal} {\bibinfo  {journal} {Phys. Rev. B}\ }\textbf {\bibinfo
  {volume} {55}},\ \bibinfo {pages} {735} (\bibinfo {year} {1997})}\BibitemShut
  {NoStop}%
\bibitem [{\citenamefont {Freitas}\ \emph {et~al.}(2016)\citenamefont
  {Freitas}, \citenamefont {Asta},\ and\ \citenamefont
  {de~Koning}}]{Freitas2016}%
  \BibitemOpen
  \bibfield  {author} {\bibinfo {author} {\bibfnamefont {R.}~\bibnamefont
  {Freitas}}, \bibinfo {author} {\bibfnamefont {M.}~\bibnamefont {Asta}}, \
  and\ \bibinfo {author} {\bibfnamefont {M.}~\bibnamefont {de~Koning}},\ }\href
  {http://www.sciencedirect.com/science/article/pii/S0927025615007089}
  {\bibfield  {journal} {\bibinfo  {journal} {Comp. Mat. Sc.}\ }\textbf
  {\bibinfo {volume} {112, Part A}},\ \bibinfo {pages} {333} (\bibinfo {year}
  {2016})}\BibitemShut {NoStop}%
\bibitem [{\citenamefont {de~Koning}(2005)}]{deKoning2005}%
  \BibitemOpen
  \bibfield  {author} {\bibinfo {author} {\bibfnamefont {M.}~\bibnamefont
  {de~Koning}},\ }\href {http://dx.doi.org/10.1063/1.1860556} {\bibfield
  {journal} {\bibinfo  {journal} {J. Chem. Phys.}\ }\textbf {\bibinfo {volume}
  {122}},\ \bibinfo {pages} {104106} (\bibinfo {year} {2005})}\BibitemShut
  {NoStop}%
\end{thebibliography}


\end{document}